\documentclass[12pt]{iopart}

\expandafter\let\csname equation*\endcsname\relax
\expandafter\let\csname endequation*\endcsname\relax
\usepackage{amsmath}
\usepackage{amssymb}
\usepackage{iopams} 
\usepackage{graphicx}  
\usepackage{subcaption} 
\usepackage{bbm} 
\graphicspath{{pic/}}  
\usepackage{hyperref}
\makeatletter
\newcommand*{\rom}[1]{\expandafter\@slowromancap\romannumeral #1@}
\makeatother
\usepackage{comment}
\usepackage{xcolor}

\usepackage{array}

\usepackage{ulem}

\definecolor{dgreen}{rgb}{0.,0.6,0.}
\definecolor{dorange}{cmyk}{0,0.6,1,0}
\definecolor{ZARMblue}{RGB}{104,153,174}

\usepackage{tikz}
\usepackage{tikz-3dplot}

\begin{document}

	\title[General orbital perturbation theory in Schwarzschild space-time]{General orbital perturbation theory in Schwarzschild space-time}

	\author{
Oleksii Yanchyshen$^{1,2}$,
Eva Hackmann$^{1}$,
Claus L\"ammerzahl$^{1}$
}

\address{$^1$ ZARM, University of Bremen, Am Fallturm 2, 28359 Bremen, Germany}
\address{$^2$ School of Mathematics and Statistics, University College Dublin, Belfield, Dublin 4, Ireland}

	\ead{oleksij.yanchyshen@gmail.com, eva.hackmann@zarm.uni-bremen.de, claus.laemmerzahl@zarm.uni-bremen.de
    }
	
	\vspace{10pt}
	\begin{indented}
		\item[]\today
	\end{indented}
	
	\bibliographystyle{alpha}
	
	\begin{abstract}
		We derive general relativistic Gaussian equations for osculating elements for orbits under the influence of a perturbing force without any restrictions in an underlying Schwarzschild space-time.  Such a formulation provides a way to describe the evolution of orbital parameters in strong gravity relativistic settings. As examples of external forces we considered Kerr and $q$-metric space-times generated forces, for which we solve  equations for osculating elements in linear approximation. For the Kerr space-time in the post-Newtonian limit, our result reproduces the well-known Lense--Thirring precession of the longitude of the ascending node.
	\end{abstract}
	
	%
	\vspace{2pc}
	
	%
	%
	%
	
	%

	\section{Introduction}
	

The direct observation of gravitational waves (GWs) by LIGO and Virgo opened a new era in the study of black holes (BHs), neutron stars, and strong‑field gravity \cite{Abbott2016Observation}. Planned space missions such as the Laser Interferometer Space Antenna (LISA) will extend GW astronomy to millihertz frequencies, enabling observations of supermassive black hole mergers and extreme mass‑ratio inspirals (EMRIs) \cite{arun2022new,AmaroSeoane2018LRR}. In particular, EMRIs probe highly relativistic orbits close to the event horizon and offer the opportunity to investigate the strong gravity regime, to explore the properties of neutron stars, and to test General Relativity (GR) and possible deviations from it. Realizing this requires us to capture the dynamics of bound orbits in the strong field with high accuracy and reasonable computational cost \cite{bambi2022handbook,BarackPound2019LRR}.

	
Because EMRI waveforms accumulate millions of radians of phase, small trajectory errors quickly become observationally significant \cite{AmaroSeoane2018LRR}. Multiple approximation schemes contribute to orbit and waveform modeling, including post‑Newtonian (PN) theory \cite{Blanchet2014LRR}, black hole perturbation theory and self‑force methods \cite{BarackPound2019LRR, pound2022black}, and fast “kludge” waveform models designed for rapid evaluation \cite{katz2021fast,Chua2017AAK,BarackCutler2004}. In this work, we develop a perturbative technique that retains the analytical spirit and usability of classical Gaussian equations while being inherently general relativistic. Our framework is designed to complement previous osculating element approaches \cite{PP2007, warburton2017evolution} and to provide a natural building block for fast orbit and kludge waveform models.

We consider generic timelike motion of a test body in a Schwarzschild spacetime subject to perturbing forces, which may arise from a deformed metric e.g., from spin or a quadrupole deformation, or from non‑gravitational interactions e.g., with a medium. Building on our previous work on general‑relativistic Gaussian perturbation equations \cite{yanchyshen2024gaussian}, we derive and organize a closed system of evolution equations for a set of seven osculating elements: energy and orbital angular momentum, complemented by inclination, complex argument of pericentre, longitude of the ascending node, as well as coordinate and proper anomaly. In contrast to formulations based on the Darwin (sometimes called Chandrasekhar) parametrization \cite{PP2007, gair2011forced, warburton2017evolution}, our evolution equations for the inclination and for the node closely mirror their familiar Newtonian counterparts, which simplifies interpretation and facilitates connection to Post-Newtonian approaches and celestial mechanics intuition. A key advantage of our approach is that the resulting evolution equations admit approximate analytical solutions, yielding efficient and computationally fast formulas suitable for practical applications. With our choice of osculating elements and parametrization, the same set of equations describes bound, unbound, and plunging trajectories, allowing for a smooth transition between bound and unbound motion. The transition from bound orbits to plunge trajectories, however, typically involves non-smooth behavior of certain parameters and, therefore, requires additional considerations\footnote{ We postpone the detailed discussion of transitions between different types of orbits to a later work. Here we focus on bound orbits.}.  Finally, when the perturbing forces retain appropriate Killing symmetries, such as stationarity and axisymmetry, our equations properly reduce to constant energy and angular momentum, as expected. 

    The main results of this paper are: (i) A general‑relativistic Gaussian perturbation framework for generic timelike Schwarzschild geodesics subject to general external (gravitational or non-gravitational) forces, formulated in terms of seven osculating elements with Newtonian‑like equations for inclination and ascending node. (ii) Compact expressions for secular rates and oscillatory corrections, with integrals reduced to combinations of elliptic and trigonometric functions that admit efficient numerical evaluation and useful analytic approximations. (iii) Validation in two representative settings: (a) the linear‑in‑spin limit of Kerr, and (b) the linear‑in‑quadrupole q‑metric. For both, we compute the nodal precession per radial period and compare against PN expansions and, in the Kerr case, we in addition calculate the pericentre advance and compare both expressions against the exact analytical solution. (iv) Demonstration that linearized solutions track the strong‑field behavior of exact Kerr geodesics more accurately than PN predictions for high‑eccentricity orbits near the horizon, while agreeing with PN in the weak‑field limit.
	
	

    We emphasize two practical aspects. First, although some terms involve products of elliptic and trigonometric functions that are cumbersome or even impossible to integrate analytically, they are straightforward to evaluate numerically at high precision, and analytic approximations for weak perturbing forces can be derived. Second, our perturbation scheme is well suited for extensions, including spinning or deformed secondaries in Schwarzschild \cite{2024PhRvL.132q1401W,2025PhRvL.134q1401S}, simple self‑force models combined with osculating elements \cite{van2018fast, warburton2012evolution}, and motion in weakly dissipative media such as plasma \cite{rezzolla2013relativistic}. A generalization to photon orbits is also natural and would be valuable for studies of BH shadows in non‑vacuum or modified‑gravity environments \cite{PerlickTsupko2022, cunha2018shadows}.
	
	As an application, we solve our perturbation equations for two small deformations of Schwarzschild: the forces induced by a small spin or a small quadrupole of the primary object. To derive the perturbing forces, we rearrange the geodesic equation in the Kerr spacetime to linear order in the spin, and the q‑metric \cite{quevedo2016quadrupolar} to linear order in the quadrupole. We compute the shifts of the inclination and of the longitude of the ascending node per radial period and compare with (i) PN asymptotics in the weak field \cite{lense1918einfluss} and (ii) the exact solution for Kerr geodesics \cite{fujita2009analytical,hackmann2012observables,cieslik2023kerr,druart2023motion}. Our linearized results agree with the exact Kerr precessions across a wide range of eccentricities and semi‑latus recta and show significantly smaller errors than first-order PN results for orbits deep in the strong field, including highly eccentric cases approaching the separatrix. In the weak‑field limit all approaches coincide, as expected.

	The remainder of the paper is organized as follows. Section \ref{geodesy::zero} reviews timelike Schwarzschild geodesics, which serve as the zeroth‑order solutions, and defines the osculating elements. Section \ref{geodesy::equations} derives the general‑relativistic Gaussian perturbation equations for timelike orbits. Section \ref{solv::lin} presents our solution strategy in the regime of weak perturbations. Section \ref{Kerr::sol} applies the method to the linear‑in‑spin Kerr case and compares pericentre and nodal precessions with PN and exact results. Section \ref{qmetr::sol} treats the linear‑in‑quadrupole q‑metric. Section \ref{disc::conc} discusses implications, limitations, and extensions. Throughout we use geometric units with $G = c = 1$ and measure all angular variables in radians.
    
	
	
	
	\section{Geodesics in the Schwarzschild space-time } \label{geodesy::zero}
	In this section we review some relevant results on geodesics in Schwarzschild spacetimes and coordinate rotations.
    The motion of a test particle in GR is defined by the geodesic equations
	\begin{equation}
		\frac{d^2 x^i}{d s^2} + \Gamma_{kl}^{i} \frac{dx^k}{ds} \frac{dx^l}{ds} =0,\label{eqv::geod}
	\end{equation}
	where $ \Gamma_{kl}^{i} $ 
    are the Christoffel symbols, $g_{ij}$ is the space-time metric with signature $(+,-,-,-)$,  $s$ is the proper time,
    and the indices run from 0 to 3. The Schwarzschild metric in spherical coordinates $(t,r,\theta,\phi)$ is given by
	\begin{equation}
		g_{tt} = 1-\frac{r_g}{r}\, , \quad 
		g_{rr} = -  \frac{1}{1-\frac{r_g}{r}}\, , \quad
		g_{\theta \theta}= - r^2 \, , \quad
		g_{\phi \phi} = - r^2 \sin^2 \theta \, . 
	\end{equation}
	Here $r_g = 2m$ is the gravitational radius and $m$ the mass. 

	It is convenient to firstly integrate once equations  (\ref{eqv::geod}) for $\phi$ and $t$ as 
	\begin{align}
		\dot{\phi} & =  L_z \frac{1}{r^2 \sin^2\theta} \,,\label{eqv::phi_s_c1}\\
		\dot{t} & = E \frac{1}{1-\frac{r_g}{r}} \, ,\label{eqv::t_s_c2}
	\end{align}
	where $L_z$ and $E$ are constants of integration, that are related to the energy and the angular momentum along the $z$-axis, and an overdot denotes a derivative with respect to proper time $s$. In the following sections, we will treat these constants as osculating elements.
    
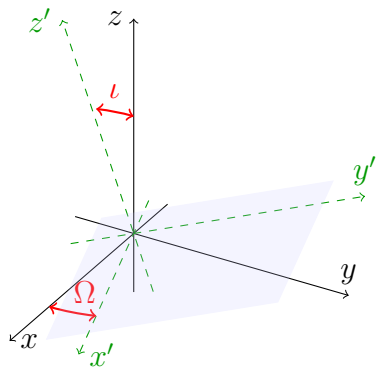
\begin{figure}
  \centering
\tdplotsetmaincoords{60}{120}
\begin{tikzpicture}[scale=3,tdplot_main_coords]
  \pgfmathsetmacro{\i}{0.3}    
  \pgfmathsetmacro{\om}{0.3}   
  \pgfmathsetmacro{\com}{cos(\om r)}   
  \pgfmathsetmacro{\som}{sin(\om r)} 
   \pgfmathsetmacro{\coi}{cos(\i r)}   
  \pgfmathsetmacro{\soi}{sin(\i r)} 

  \draw[->] (-0.3,0,0) -- (1.1,0,0) node[right]{$x$};
  \draw[->] (0,-0.3,0) -- (0,1.1,0) node[above]{$y$};
  \draw[->] (0,0,-0.3) -- (0,0,1.1) node[left]{$z$};
  
  \draw[->,dashed,dgreen] (-0.3*\com, -0.3*\som, 0.) -- (1.1*\com, 1.1*\som, 0.) node[right]{$x'$};
  \draw[->,dashed,dgreen] (0.3* \coi *\som, -0.3* \coi*\com, -0.3* \soi) -- (-1.1* \coi *\som, 1.1* \coi*\com, 1.1* \soi) node[above]{$y'$};
  \draw[->,dashed,dgreen] (-0.3* \soi*\som, 0.3* \com* \soi, -0.3*\coi) -- (1.1* \soi*\som, -1.1* \com* \soi, 1.1*\coi) node[left]{$z'$};
  \draw[<->,red,thick] (0.75,0,0) arc[start angle=0, end angle=\om*180/3.1416, radius=0.75];
  \node[red] at (0.59*\com,0.25*\som,0) {$\Omega$};
  
  \draw[<->,red,thick] 
    (0,0,0.6) 
    arc[start angle=0, end angle=\i*180/3.1416, radius=-0.6]; 
\node[red] at (0,-.1,0.7) {$\iota$}; 

  \pgfmathsetmacro{\px}{.9}  
  \pgfmathsetmacro{\nx}{0.2}  
  \pgfmathsetmacro{\py}{.9}  
  \pgfmathsetmacro{\ny}{0.2}  
  \pgfmathsetmacro{\vx}{\com}                 
  \pgfmathsetmacro{\vy}{\som}                 
  \pgfmathsetmacro{\vz}{0}                    
  \pgfmathsetmacro{\ux}{- \coi * \som}        
  \pgfmathsetmacro{\uy}{  \coi * \com}        
  \pgfmathsetmacro{\uz}{  \soi}               

  \pgfmathsetmacro{\VAx}{- \nx*\vx - \ny*\ux}
  \pgfmathsetmacro{\VAy}{- \nx*\vy - \ny*\uy}
  \pgfmathsetmacro{\VAz}{- \nx*\vz - \ny*\uz}

  \pgfmathsetmacro{\VBx}{  \px*\vx - \ny*\ux}
  \pgfmathsetmacro{\VBy}{  \px*\vy - \ny*\uy}
  \pgfmathsetmacro{\VBz}{  \px*\vz - \ny*\uz}

  \pgfmathsetmacro{\VCx}{  \px*\vx + \py*\ux}
  \pgfmathsetmacro{\VCy}{  \px*\vy + \py*\uy}
  \pgfmathsetmacro{\VCz}{  \px*\vz + \py*\uz}

  \pgfmathsetmacro{\VDx}{- \nx*\vx + \py*\ux}
  \pgfmathsetmacro{\VDy}{- \nx*\vy + \py*\uy}
  \pgfmathsetmacro{\VDz}{- \nx*\vz + \py*\uz}

  \coordinate (V1) at (\VAx,\VAy,\VAz);
  \coordinate (V2) at (\VBx,\VBy,\VBz);
  \coordinate (V3) at (\VCx,\VCy,\VCz);
  \coordinate (V4) at (\VDx,\VDy,\VDz);

  \filldraw[blue!20,opacity=0.2] (V1) -- (V2) -- (V3) -- (V4) -- cycle;
   
  \end{tikzpicture}
  \caption{Two coordinate systems: the general system $\vec{x}$ and the system $\vec{x}'$ in which the orbital plane lies in the equatorial plane, connected by the rotation (\ref{changing_of_the_variables}).}
\label{plot::chang:variables}
\end{figure}

    Due to spherical symmetry, the motion of the test particles is confined to an orbital plane. We use a
    rotation $A \vec{x}'= \vec{x}$ that connects our coordinate system  $\vec{x}= ( r \cos \phi \sin \theta, r \sin \phi  \sin \theta , r \cos \theta)$ with the system $\vec{x}'=( r \cos \varphi \sin \vartheta, r \sin \varphi  \sin \vartheta , r \cos \vartheta)$, where the orbital plane is $\vartheta=\frac{\pi}{2}$, as illustrated in figure~\ref{plot::chang:variables},
	\begin{eqnarray}
		A = \left(
		\begin{array}{ccc}
			\cos \Omega  & -\cos \iota  \sin \Omega  & \sin \iota  \sin \Omega  \\
			\sin \Omega  & \cos \iota  \cos \Omega  & -\sin \iota \cos \Omega  \\
			0 & \sin \iota  & \cos \iota  \\
		\end{array}
		\right)\,. \label{changing_of_the_variables}
	\end{eqnarray}
	We then find
	\begin{eqnarray}
		&\cos\theta =\sin \iota  \sin \varphi,\label{eqv::cos_thet_varphi_0}\\
		&\tan (\phi-\Omega) =\cos\iota \tan\varphi,\label{eqv::phi_varphi_0}
	\end{eqnarray}
	where $\iota$ is orbital inclination and $\Omega$ is longitude of the ascending node. 
    Eliminating $ \varphi$ from  (\ref{eqv::cos_thet_varphi_0})-(\ref{eqv::phi_varphi_0}) we find after simplification 
	\begin{eqnarray}
		&\cos^2 \theta =\frac{\tan^2 (\Omega -\phi)}{\cot ^2\iota +\csc ^2\iota \tan ^2(\Omega -\phi)}\label{eqv::theta_phi_0}.
	\end{eqnarray}
	By taking the derivative of (\ref{eqv::theta_phi_0}) with respect to proper time, and using (\ref{eqv::phi_s_c1}) to eliminate $\dot{\phi}$, we obtain
	\begin{eqnarray}
		&\dot{\theta}= -\frac{ L_z \tan \iota  \cos (\phi -\Omega )}{r^2}   ,\label{eqv::dot_theta_phi}
	\end{eqnarray}
	which satisfies equation (\ref{eqv::geod}) for $\theta$.  
	
The solution of the geodesic equations (\ref{eqv::geod}) within the orbital plane can be written in many different ways, see e.g. \cite{1917KNAB...19..197D,doi:10.1098/rspa.1920.0019,Hagihara,fc643266-b6ab-3a0d-ae6a-ccbfd43fdadb,Scharf2011,kostic2012analytical}. 
 We use here the solution in terms of Weierstrass elliptic function $\wp$, as in \cite{Hagihara}
	\begin{eqnarray}
		r = \frac{ r_g }{ \wp(\frac{\varphi-\varphi_0}{2} + \omega_2) + \frac{1}{3}}\,, \label{eq:solr}
	\end{eqnarray}
	where $\varphi_0$ is the argument of the periastron (which we will also treat together with $\omega_2$ as an osculating element in the next section), and $\omega_2= \omega_1+\omega_3$, where $\omega_1$ is a real half-period and $\omega_3$ is a complex  half-period of the Weierstrass function $\wp$. The shift by $\omega_3$ is required to ensure that (\ref{eq:solr}) describes a bound orbit or an unbound scattering orbit \cite{lawden2013elliptic}. In the case of plunging orbits, the same expression for $r$ applies, but without the half-period shift $\omega_3$ in the argument of the Weierstrass function. For bound orbits, the shift by $\omega_2$ is required to ensure that $\varphi_0$ is indeed the argument of periastron. Invariants $g_2$ and $g_3$ of $\wp $, which we omitted for brevity in eq.~\eqref{eq:solr}, serve as geometrical characteristics of the trajectory and are related to the angular momentum $L_z = L \cos \iota $ and the energy $E$ as
	\begin{eqnarray}
		g_2 = 4  \left( \frac{1}{3}-\frac{ r_g^2}{L^2}\right),\label{conection::g2_c1}\\
		g_3 = \frac{4}{3} \left( \frac{2}{9}-\frac{ \left(3 E^2-2\right) r_g^2}{ L^2}\right)\label{conection::g3_c2}.
	\end{eqnarray}
In \ref{appendix::newtonian::limit} we present the Newtonian limit of the solution (\ref{eq:solr}) together with the invariants of Weierstrass elliptic function $g_2$ and $g_3$. For more details on the solution in the equatorial plane coordinate system, we refer to our previous paper \cite{yanchyshen2024gaussian} or to the original results of Hagihara \cite{Hagihara}.

	\section{General relativistic orbital perturbations}\label{geodesy::equations}

\subsection{Evolution equations for orbital elements}
    
The motion of a test particle in GR under the influence of a general external force is given by 
	\begin{eqnarray}
		\frac{d^2 x^i}{d s^2} + \Gamma_{kl}^{i} \frac{dx^k}{ds} \frac{dx^l}{ds} =f^i, \label{eq::geod::with:force}
	\end{eqnarray}
where the force satisfies the orthogonality relation
	\begin{eqnarray}
		f^i \dot{x}_i =0, \label{eqv::ortogonality::f}
	\end{eqnarray}
that follows from the normalisation condition $\dot{x}_i \dot{x}^i=1$ (here and throughout the article we consider only timelike trajectories). To solve the four second order differential equations \eqref{eq::geod::with:force}, we generally need to choose eight initial values or, equivalently, eight contant orbital elements. However, the constraint \eqref{eqv::ortogonality::f} removes one of these, leaving us with seven independent orbital elements. The choice of these elements is arbitrary and can be adjusted to produce convenient equations. In the following, we denote our set of seven orbital elements by $I_A$.

To formulate the perturbation equations we use a technique inspired by the method of variation of constants, which is widely used in Newtonian gravity (for example, see \cite{Klioner, soffel2019applied}).  First, we assume that the perturbed orbit at each instant of proper time coincides with an instantaneous geodesic. By going smoothly from one instantaneous geodesic to the next, we endow the orbital elements with a time dependence. To close the system of equations, we then require the osculating condition, 
	\begin{eqnarray}
		&\sum_{A} \frac{\partial z^\alpha}{\partial I_A} \dot{I}_A =0,  \label{formal:osculation::cond}
	\end{eqnarray}
	where $z^\alpha = (t,r,\phi, \theta)$. The evolution equations then read
	\begin{eqnarray}
    \quad \sum_{A} \frac{\partial \dot{z}^\alpha}{\partial I_A} \dot{I}_A =f^\alpha \,.
    \end{eqnarray}
For the Schwarzschild metric with an external force the equations of motion are
	\begin{eqnarray}
		\ddot{t}  +\frac{   r_g }{r^2(1- r_g/r)} \dot{t} \dot{r} =f^t,\label{eqv::t_s_f}\\
		\ddot{r} +\dot{t}^2\frac{ \left(1-r_g/r\right) r_g}{2 r^2}-\dot{r}^2 \frac{ r_g}{2 r^2}\frac{1}{1-  r_g/r}-\left(r- r_g \right) \left(\dot{\phi}^2  \sin ^2 \theta +\dot{\theta}^2 \right)=f^r,\label{eqv::r_s_fr}\\
		\ddot{ \theta}-\dot{ \phi}^2 \sin\theta  \cos\theta +\frac{2  }{r} \dot{ \theta} \dot{r}=f^\theta,\label{eqv::theta_s_f}\\
		\ddot{ \phi}+2 \dot{ \phi}  \dot{ \theta} \cot \theta  +\frac{2  }{r} \dot{ \phi} \dot{ r} =f^\phi. \label{eqv::phi_s_f}
	\end{eqnarray}
Note that it is also possible to decompose the force components into the STW coordinate system, as is usually done in Newtonian theory \cite{Klioner}; however, we will omit this here. In the following, we will use the relations (\ref{eqv::phi_s_c1}), (\ref{eqv::t_s_c2}) 
for $\dot{\phi}$ and $\dot{t}$ 
	together with the connection between $r$ and the argument of periastron $\varphi_0$ \eqref{eq:solr} and its derivative,
	\begin{eqnarray}
		&\dot{r} = -\frac{L}{2 r_g } \wp'\left(\frac{\varphi- \varphi_0}{2} +\omega_2 \right)\,.\label{eqv::dot_r_varphi} 
	\end{eqnarray}
	We also use (\ref{eqv::theta_phi_0}) and (\ref{eqv::dot_theta_phi}) for the relations of $\phi$, $\theta$, and $\dot{\theta}$ with the ascending node $\Omega$ and the inclination $\iota$.
    
	The osculating elements, which were constant in the unperturbed case, are now functions of time. Our aim is to derive evolution equations for these functions that, while satisfying the relations above, also satisfy the non-homogeneous equations (\ref{eqv::t_s_f})–(\ref{eqv::phi_s_f}). 
\paragraph{Energy and angular momentum:}	By substituting the relations for $\dot{\phi}$ and $\dot{t}$ (\ref{eqv::phi_s_c1}) and (\ref{eqv::t_s_c2})  into geodesic equations  (\ref{eqv::phi_s_f}) and (\ref{eqv::t_s_f})  we obtain the following evolution equations for the osculating elements $L_z$ and $E$ 
	\begin{eqnarray}
		\dot{L}_z =  r^2 \sin^2\theta f^{\phi} ,\label{eqv::dot_c3}\\ 
		\dot{E}  = \left(1- \frac{r_g}{r}\right) f^t. \label{eqv::dot_c2} 
	\end{eqnarray}
\paragraph{Inclination and ascending node:}	Similarly, by substituting relation  (\ref{eqv::dot_theta_phi}) for $\dot{\theta}$ into the geodesic equation (\ref{eqv::theta_s_f}), and using relation (\ref{eqv::theta_phi_0}) between $\dot{\iota}$ and $\dot{\Omega}$  together with equation (\ref{eqv::dot_c3}), we obtain, after some simplifications, the evolution equations for the inclination $\iota$ and  the ascending node $\Omega$ in a manner analogous to the non-relativistic case \cite{Klioner,soffel2019applied}:
	\begin{eqnarray}
		\dot{\iota} = -\frac{ r^2 \cot \varphi  \sin ( \Omega -\phi)}{L} f^\theta+\frac{  r^2 \sin \iota  \cos ^2\varphi }{L} f^\phi \label{eq::diota},\\
		\dot{\Omega} = \frac{ r^2 \sin (\Omega -\phi)}{L \sin\iota } f^{\theta} -\frac{ r^2 \sin \varphi  \cos\varphi}{L} f^\phi  \label{eq::domega}.
	\end{eqnarray}
    
\paragraph{Complex argument of periastron:}   Finally, for non-plunging orbits it is useful to define the following (complex-valued) osculating element $\bar{\varphi}_0(s)$ as 
\begin{eqnarray}
		\bar{\varphi}_0(s)= 2\omega_2(s)- \varphi_0(s)\,.
\end{eqnarray}
This definition allows to integrate the evolution of the half-periods $\omega_2=\omega_1+\omega_3$ into the argument of periastron instead of treating it separately, which will simplify calculations significantly. We can derive the evolution equation of $\bar{\varphi}_0$ by substituting expression (\ref{eq:solr}) for $r$  into relation (\ref{eqv::dot_r_varphi})  for $\dot{r}$  and by using exact expressions for $\frac{\partial \wp}{\partial g_2}$ and $\frac{\partial \wp}{\partial g_3}$ \cite{FunctionsWolfram}. After simplification, and using equations (\ref{eqv::dot_c3})-(\ref{eq::diota}) for $L_z$, $E$ and $\iota$, the evolution equation for $\bar{\varphi}_0$ can be written as 
	\begin{eqnarray}
		\dot{\bar{\varphi}}_0  =  A_t f^t + A_\phi (1+\sin ^2\iota \cos 2 \varphi )  f^\phi +A_\theta \sin \iota \cos (\Omega-\phi)  f^\theta,\label{eqv::dot_phib0_A}
	\end{eqnarray}
	where the explicit expressions for $A_t$, $A_\phi$ and $A_\theta$ are given in \ref{exact::expressions::eq}. In general the evolution equation (\ref{eqv::dot_phib0_A}) becomes ill-defined because of terms proportional to  $\frac{1}{\wp'}$ and therefore in general requires regularization. 
	However, using  the orthogonality relation (\ref{eqv::ortogonality::f}) for the force, we can rewrite  evolution equation (\ref{eqv::dot_phib0_A}) in terms of $f^\phi$, $f^r$ and $f^\theta$. By expressing all cubic and higher powers of $\wp$ in terms of $\wp'^2$, the problematic factors of $\wp'$ in the denominator cancel out. The resulting expression takes the form
    \begin{eqnarray}
		\dot{\bar{\varphi}}_0  =  B_r f^r + \cos \iota B_\phi     f^\phi +B_\theta  \cos (\Omega-\phi)  f^\theta,\label{eqv::dot_phib0}
	\end{eqnarray}
 where the explicit expressions for $B_r$, $B_\phi$ and $B_\theta$ are given in \ref{exact::expressions::eq}. This rewriting completely removes the singular 
$\frac{1}{\wp'}$  terms, and therefore eliminates the need for any regularization.

\paragraph{Coordinate and proper anomaly:}  
In our previous work \cite{yanchyshen2024gaussian}, we omitted the explicit equations for $s_0$ and $t_0$ due to their complexity, noting that the evolution of $s$ and $t$ can be calculated directly from equations (\ref{eqv::phi_s_c1})-(\ref{eqv::t_s_c2}), or equivalently, from the corresponding definitions of the anomalies. Here we close this gap and define the general relativistic coordinate anomaly $M_t$ as 
 \begin{eqnarray}
	M_t := \frac{L}{E r_g^2} (t-t_0) = \int  \frac{d\varphi}{ \left(\frac{1}{3}+\wp(v)\right)^2 \left(\frac{2}{3}-\wp(v)\right)} , \label{eq::mean::def} 
\end{eqnarray}
where we used equation (\ref{eqv::t_s_c2}) for the last equality. In the appropriate limit, this definition reduces to a quantity proportional to the Keplerian mean anomaly. Taking the derivative of both sides of the definition (\ref{eq::mean::def}), we obtain the evolution equation in the form
\begin{eqnarray}
		 M'_t = \frac{1}{ \left(\frac{1}{3}+\wp(v)\right)^2 \left(\frac{2}{3}-\wp(v)\right)}, 
         \label{eq:evolutionMt}
\end{eqnarray}
where $v = \frac{1}{2} (\varphi+\bar\varphi_0)$ and the prime denotes differentiation with respect to $\varphi$.

Analogously, we define the general relativistic proper anomaly $M_s$ as 
\begin{equation}
M_s := \frac{L}{r_g^2} (s-s_0) = \int \frac{d\varphi}{ \left(\frac{1}{3}+\wp(v)\right)^2 } \label{eqv::s_intdph}\,, 
\end{equation}
and find the corresponding evolution equation in the form 
\begin{eqnarray}
		M'_s = \frac{1}{ \left(\frac{1}{3}+\wp(v(s), g_2(s), g_3(s))\right)^2 } .
        \label{eq:evolutionMs}
\end{eqnarray}
 We choose to write the evolution equations for anomalies $M_s$ and $M_t$ in terms of the angular variable $\varphi$ rather than the proper time. This choice is motivated by two reasons. First, it simplifies the analysis: expressing $M_s$ in terms of the proper time would require differentiating the Weierstrass elliptic functions $\zeta(v(s), g_2(s), g_3(s))$ and $\sigma(v(s), g_2(s), g_3(s))$ with respect to all their arguments, which we prefer to avoid. Second, and more importantly, within our perturbation framework all quantities are naturally parametrized by $\varphi$, making this choice both consistent with and preferable for the overall formalism.

 The evolution equations \eqref{eqv::dot_c3},   \eqref{eqv::dot_c2}, \eqref{eq::diota}, \eqref{eq::domega}, \eqref{eqv::dot_phib0}, \eqref{eq:evolutionMt}, \eqref{eq:evolutionMs} for the seven osculating elements are the first main result of this paper. They form a complete general‑relativistic Gaussian perturbation framework for generic timelike Schwarzschild geodesics subject to general external (gravitational or non-gravitational) forces.

	\subsection{Linearization}\label{solv::lin}
	The evolution equations (\ref{eqv::dot_c3})- \ref{eq::domega}, \ref{eqv::dot_phib0}) form a set of nonlinear first order differential equations, which generally cannot be solved analytically for an arbitrary force. We can symbolically rewrite the equations as
	\begin{eqnarray}
		& \dot{I}_A(s)= F_{A}(\{I_A, z^j,f^k\}),\label{eq::I::s::F} 
	\end{eqnarray}
	 where $F_{A}(\{I_A, z^j,f^k\})$ denote the RHSs of these equations. In the case where the force is proportional to a small parameter $\lambda$, we can  expand the equations  (\ref{eq::I::s::F}) to linear order
	\begin{eqnarray}
		& \dot{I}_A(s)=  F_{A}(\{I_A, z^j,\lambda\}) +O(\lambda^2) \label{eq::I::s::F_const} 
	\end{eqnarray}
	where $E$, $L_{z}$, $\iota$, $\Omega$ and $ \bar{\varphi}_0$ on the RHSs are now constants. Thus, for a force that does not depend explicitly on proper time $s$, the RHSs depend only on $\varphi(s)$. This allows us to reparametrise them in terms of  $\varphi$ 
	\begin{eqnarray}
		& I'_A(\varphi)= F_{A}(E,L_z,\iota,\Omega, \bar{\varphi}_0,\varphi , \lambda),
	\end{eqnarray}
	enabling us to (at least in principle)  analytically integrate these equations.

For calculations involving the anomalies $M_t$ and $M_s$, we can use the following linearization scheme: In the osculating elements framework, the parameters $g_2$, $g_3$, and $\bar \varphi_0$ are functions of $\varphi$. We expand them in series with respect to a small parameter $\lambda$, $g_2=g_2(0)+ \lambda  \delta g_2 $, $g_3=g_3(0)+ \lambda \delta g_3$, and $\bar{\varphi}_0 =\bar{\varphi}_0(0)+ \lambda \delta \varphi_0$. Accordingly, expanding the anomaly $M_t$ (\ref{eq::mean::def}) in a series in $\lambda$ gives
\begin{eqnarray}
M_t = M_t^0 + \lambda M_t^{g_2} + \lambda M_t^{g_3} + \lambda M_t^{\varphi_0},
\end{eqnarray}
with 
\begin{eqnarray}
		  M_{t}^{g_2}=   \int  d\varphi  \frac{ \delta g_2   (3 \wp -1) \left(-36 g_3 \left(\zeta \wp'+2 \wp ^2\right)+2 g_2^2 \left(   v   \wp' +2 \wp \right)+12 g_3 g_2\right)}{8 \left(g_2^3-27 g_3^2\right) (\frac{2}{3}- \wp )^2 ( \frac{1}{3}+\wp )^3}, \nonumber \\ 
          M_{t}^{g_3} = -   \int  d\varphi \frac{   \delta g_3   (3 \wp -1) \left(-12 g_2 \left(\zeta  \wp' +2 \wp ^2\right)+18 g_3 \left(  v \wp'+2 \wp \right)+4 g_2^2\right)}{4 \left(g_2^3-27 g_3^2\right) (\frac{2}{3}-3 \wp )^2 ( \frac{1}{3}+\wp )^3},  \nonumber \\ 
          M_{t}^{\varphi_0} =  \int  d\varphi \frac{  \delta \bar{\varphi}_0   (3 \wp -1) \wp' }{2 (\frac{2}{3}- \wp )^2 (\frac{1}{3}+\wp )^3},
\end{eqnarray}
where we omitted the argument of $\wp(v)$ and $\zeta(v)$ for simplicity.

Using the same approach to the evolution 
of the proper anomaly $M_s$ (\ref{eqv::s_intdph}) we have
\begin{eqnarray}
M_s = M_s^0 + \lambda M_s^{g_2} + \lambda M_s^{g_3} + \lambda M_s^{\varphi_0}.
\end{eqnarray}
with 
\begin{eqnarray}
		  M_s^{g_2}=   -\int  d\varphi    \frac{  \delta g_2   \left(-36 g_3 \left(\zeta   \wp'+2 \wp ^2\right)+g_2^2 (2 v  \wp'+4 \wp )+12 g_3 g_2\right)}{4 \left(g_2^3-27 g_3^2\right) (\frac{1}{3}+\wp )^3}, \nonumber \\ 
          M_s^{g_3} =   \int  d\varphi \frac{  \delta g_3   \left(-12 g_2 \left(\zeta  \wp' +2 \wp ^2\right)+9 g_3 (2 v \wp' +4 \wp )+4 g_2^2\right)}{2 \left(g_2^3-27 g_3^2\right) (\frac{1}{3}+\wp)^3},  \nonumber \\ 
          M_s^{\varphi_0} =  -\int  d\varphi \frac{  \delta \bar{\varphi}_0    \wp' }{(\frac{1}{3}+\wp )^3},
\end{eqnarray}
where we again omitted the argument of $\wp(v)$ and $\zeta(v)$ for simplicity.

\subsection{Secular motion}	
From an observational point of view, the most interesting perturbations are those that accumulate over time. Therefore, we are interested in these long-term secular effects of the perturbing force. In order to obtain these, we may expand eq.~\eqref{eq::I::s::F} in a proper or a coordinate time Fourier series. For the former, we find 
	\begin{eqnarray}
		& \dot{I}_A(s)= F_{A, 0} + \sum_{k\ne 0}^{\infty}  F_{A,k} e^{ i k M_{s} } , 
	\end{eqnarray}
	where $k =\frac{ 2\pi n }{M_s(4 \omega_1) }$ with $n \in N $, $M_s(4 \omega_1)=: T_s $ is the proper anomalistic period, and $4 \omega_1$ is the period of the solution $r(\varphi)$ as a function of orbital plane angle $\varphi$ in equation \eqref{eq:solr}. 
	We can integrate these equations neglecting all oscillatory terms as
	\begin{eqnarray}
		& I_A(s)= I_{A}(s_0) +  (s-s_0) F_{A,0} , 
	\end{eqnarray}
	where we define the zeroth harmonics as 
	\begin{eqnarray}
		&  F_{A,0}= \frac{1}{T_s}\int_{0}^{T_s} F_{A}(s)  d M_{s}\,
	\end{eqnarray}
	and $\frac{1}{T_s}\int_{0}^{T_s} $ as the average over the proper anomalistic period $T_s$.

	We may proceed in an analogous way for the coordinate time expansion. We first symbolically write the evolution equations in coordinate time, 
	\begin{eqnarray}
		& \dot{I}_A(t)= G_{A} (\{I_A, z^j\}).
	\end{eqnarray}
	Then the coordinate time Fourier series is
	\begin{eqnarray}
		& \dot{I}_A(t)= G_{A,0} + \sum_{k\ne 0}^{\infty}  G_{A,k}  e^{ i k M_{t} } ,  
	\end{eqnarray}
	with the zeroth harmonics defined as follows
	\begin{eqnarray}
		&  G_{A,0} = \frac{1}{T_t}\int_{0}^{T_t}  G_{A}(t)   d M_{t}, 
	\end{eqnarray}
	where $\frac{1}{T_t} \int_{0}^{T_t}$ is the average over an anomalistic coordinate period $T_t=M_t(4\omega_1)$.
	Thus, we have secular perturbations  
	\begin{eqnarray}
		& I_A(t)= I_{A}(t_0) +  (t-t_0) G_{A,0} .
	\end{eqnarray}
	Also, there is a connection between proper and coordinate time secular perturbations 
	\begin{eqnarray}
		& G_{A,0} = \frac{T_s}{T_t E} F_{A,0}.\label{eq::conect:GF}
	\end{eqnarray}
It is also important to define a secular shift per revolution for osculating element $I_A$ as 
    \begin{eqnarray}
		& \Delta_{A} = I_A(s_0+s_p)-I_A(s_0) = s_p  F_{A,0}, \label{eq:def::sec:cor}
	\end{eqnarray}
   where $s_p = \frac{r_g^2}{L} T_s$ is the proper period, which we use extensively in the sections below.

	

	\section{  Orbital perturbations from gravitomagnetism  }	\label{Kerr::sol}
	In this section, we will investigate the impact of a disturbing force that models a rotation of the central object. For this, we derive the perturbation force from the geodesic equation in Kerr spacetime. In analogy with the spacetime investigated by Lense and Thirring \cite{lense1918einfluss}, which characterizes any first-order rotational frame, the corresponding perturbative force may manifest in various rotational effects associated with spacetime dragging, allowing us to model systems governed by strong gravitational fields and slow rotation. 
    We can write the Kerr metric in Boyer–Lindquist coordinates ($t, r, \theta, \phi$) as     
	\begin{eqnarray}
		&g_{tt}=  1-\frac{r_g r }{\rho^2} , \quad g_{t\phi}=-\frac{r_g a r \sin^2 \theta }{\rho^2},  \quad g_{rr}=-\frac{\rho^2}{\Delta}, \quad g_{\theta\theta}=-\rho^2,\\
		&g_{\phi\phi}=-\left(r^2 +a^2 + \frac{r_g a^2 r  \sin^2 \theta}{\rho^2} \right)\sin^2 \theta,
	\end{eqnarray}
	where $\rho^2 = r^2+ a^2 \cos^2\theta$ and $\Delta=  r^2 -r_g r+ a^2 $ with a spin parameter $a$. In order to solve the geodesic equations perturbatively we can expand them as a power series in the dimensionless spin parameter (angular momentum) $J = \frac{a}{r_g}$. As a result we obtain geodesic equations for the Schwarzschild metric with force 
	\begin{eqnarray}
		f^t =  J \frac{3  r_g^2 \dot{r} \dot{\phi}  \sin ^2\theta }{r^2 (1-\frac{r_g}{r})},\label{def::ft::kerr}\\
		f^r = J \frac{ r_g^2 \dot{t}\dot{\phi}  \sin ^2 \theta  }{r^2} \left(1-\frac{r_g}{r}\right) , \\
		f^\theta =  -J \frac{ r_g^2 \dot{t}\dot{\phi}  \sin 2 \theta }{r^3}  \label{def::fth::kerr},\\
		f^\phi = -  J  r_g^2 \left(\frac{\dot{t}\dot{r}}{r^4 (1-\frac{r_g}{r})}-\frac{2 \dot{t}\dot{\theta}  \cot \theta }{r^3}\right) \label{def::fph::kerr} .
	\end{eqnarray}
   In the following subsections, we integrate the linearised 
    evolution equations for the osculating elements $\iota$, $\Omega$, $L_z$, $E$ and $\bar{\varphi}_0$  with this force.

	\subsection{Equation of the inclination $\iota $ and the longitude of the ascending node $\Omega$ }
	By using the force components (\ref{def::fth::kerr})-(\ref{def::fph::kerr}) in equations (\ref{eq::diota})-(\ref{eq::domega}) we find for the inclination $\iota $ and the longitude of the ascending node $\Omega$ 
	\begin{eqnarray}
		\dot{\iota} = -\sin \iota \frac{J E r_g^2}{r^2} \left(\frac{  \sin 2 \varphi }{r (1-\frac{r_g}{r})}+\frac{  \dot{r}   \cos ^2\varphi }{L  (1-\frac{r_g}{r})^2} \right),\label{eq::dot_iota}\\
		\dot{\Omega}= \frac{J E r_g^2}{ L  r^2} \left(\frac{2   L   \sin ^2\varphi }{r (1-\frac{r_g}{r})}+\frac{    \dot{r} \sin \varphi  \cos \varphi }{ (1-\frac{r_g}{r})^2}\right). \label{eq::dot_Omega}
	\end{eqnarray}
	We are interested to find the long-term secular contributions of the rotation on the inclination and the node, that will generalise the post-Newtonian results for the Lense-Thirring shift. Solving secular linearised equations (\ref{eq::dot_iota}) and (\ref{eq::dot_Omega}) for $\iota$ and $\Omega$ we obtain
	\begin{eqnarray}
		\iota(s) = \iota(s_0) +  (s-s_0)   F_{\iota, 0},   \\
		\Omega(s) = \Omega(s_0) +  (s-s_0)   F_{\Omega, 0},  
	\end{eqnarray}
	where 
	\begin{eqnarray}
		F_{\iota,0} & = & -\sin \iota \frac{J E r_g^2  }{T_s} \int_0^{T_s}     \frac{1}{r^2} \left(\frac{  \sin 2 \varphi }{r (1-\frac{r_g}{r})}+\frac{  \dot{r}   \cos ^2\varphi }{L  (1-\frac{r_g}{r})^2} \right) dM_s \nonumber \\
		& = &  -\sin \iota \frac{J E   }{  T_s} \int_0^{4\omega_1} \left(\frac{  \sin 2 \varphi }{r (1-\frac{r_g}{r})}+\frac{  \dot{r}   \cos ^2\varphi }{L  (1-\frac{r_g}{r})^2} \right) d\varphi  ,   
	\end{eqnarray}
	and analogously
    \begin{eqnarray}
		&F_{\Omega, 0} = 
        \frac{ J E   }{  T_s} \int_0^{4\omega_1}    
		\frac{1}{  L} \left(\frac{2   L   \sin ^2\varphi }{r (1-\frac{r_g}{r})}+\frac{    \dot{r} \sin \varphi  \cos \varphi }{ (1-\frac{r_g}{r})^2}\right) d\varphi  ,   
	\end{eqnarray}
	which we can calculate numerically.

    The secular effect $\Delta_\Omega$ on the longitude of the node per revolution is given by (\ref{eq:def::sec:cor}), 
\begin{eqnarray}		 &\Delta_{\Omega} = \Omega(s_0+s_p) - \Omega(s_0) = s_p F_{\Omega,0}\,. \label{eq::q_metr::delta::omega::sol}  \end{eqnarray} 
From this we find
	\begin{eqnarray}
		&\Delta_{\Omega} = s_p F_{\Omega,0} = \frac{ J E r_g^2  }{L^2} \int_0^{4\omega_1}    
		\left(\frac{2   L   \sin ^2\varphi }{r (1-\frac{r_g}{r})}+\frac{    \dot{r} \sin \varphi  \cos \varphi }{ (1-\frac{r_g}{r})^2}\right) d\varphi \,, \label{eq::delta:omega}  
	\end{eqnarray} 
   where $s_p$ is the proper orbital period. The resulting values are computed numerically in table~(\ref{tab:my_label_1_plus_2}) and discussed in section \ref{sec:compKerr} below. To compare this result to the usual post-Newtonian expressions, we derive this limit from the above equation. We find
	\begin{eqnarray}
		&\Delta_{\Omega}  = \frac{2 J E r_g^2  }{L} \int_0^{2 \pi}    \frac{  \sin^2 \varphi  }{ r    } d\varphi+ \frac{ J E r_g^2  }{L^2} \int_0^{2\pi}      \dot{r} \sin \varphi  \cos \varphi  d\varphi + O\left(\frac{r_g}{r}\right). \label{eq::delta::Omega::pn}
	\end{eqnarray} 
	Using Kepler's solution $r \approx \frac{p}{1+e \cos(\varphi+\varphi_0)} $ and integrating (\ref{eq::delta::Omega::pn}) we have
	\begin{eqnarray}
		&\Delta_{\Omega}  = \frac{2 \pi  J E r_g^2}{L p} + O\left(\frac{r_g}{p}\right) . 
	\end{eqnarray} 
	We can rewrite this result by taking into account the expression for the ratio of energy $E$ and angular momentum $L$ in terms of apocenter $r_a$ and pericentre $r_p$ distances, 
	\begin{eqnarray}
		&  \frac{  E  }{L } = -\frac{\sqrt{r_a-r_g} \sqrt{r_a+r_p} \sqrt{r_p-r_g}}{r_a r_p \sqrt{r_g} } , 
	\end{eqnarray} 
	and finally we get 
	\begin{eqnarray}
		&\Delta_{\Omega}  = J \frac{2 \sqrt{2} \pi   r_g^{3/2}}{p^{3/2}} + O\left(\frac{r_g}{p}\right) . \label{eq::delta::omega::PN}
	\end{eqnarray} 
	This is the same as the shift obtained from post-Newtonian calculations, and known as Lense--Thirring effect (see for example \cite{soffel2019applied}, \cite{lense1918einfluss}, \cite{ruffini2003nonlinear}).

	The secular change of the inclination $\iota$ per revolution is
	\begin{eqnarray}
		&\Delta_{\iota}  =  -\sin \iota \frac{J E r_g^2  }{L  } \int_0^{4\omega_1} \left(\frac{  \sin 2 \varphi }{r (1-\frac{r_g}{r})}+\frac{  \dot{r}   \cos ^2\varphi }{L  (1-\frac{r_g}{r})^2} \right) d\varphi,   
	\end{eqnarray} 
	and, by calculating the post-Newtonian asymptotic in a similar manner, we obtain, owing to the periodicity of all terms,
	\begin{eqnarray}
		&\Delta_{\iota}  = O\left(\frac{r_g}{p}\right),   
	\end{eqnarray}
	which is also a known post-Newtonian result.

\begin{table}
		\centering 
       \begin{center}
\begin{tabular}{ | m{4.6cm} | m{2.1cm}| m{2.1cm} | m{2.1cm}|  m{2.1cm}|  } 
  \hline
			& $\Delta_{\Omega}    $ &  $|1-\frac{\Delta_{\Omega}}{\Delta_{\Omega}^{PN}}|$   & $|1-\frac{\Delta_{\Omega}^{FH}}{\Delta_{\Omega}^{PN}} |$  
			& $|1-\frac{\Delta_{\Omega}^{FH}}{\Delta_{\Omega}}|$ 
			\\\hline
			$r_a=2\cdot10^{ 5}$, $r_p=1.5\cdot10^{ 5}$,  ($e=0.14$, $p=~1.71~\cdot10^{ 5}$) & $ 1.2519\cdot 10^{-8} $& $1.4\cdot 10^{-6}$ & $1.2\cdot 10^{-4}$ &   $ 1.2 \cdot 10^{-4} $ \\\hline 
			$r_a=9000$, $r_p=8500$, ($e=~0.028$, $p=8743$) &  $  1.087 \cdot 10^{-6} $ &  $4.6\cdot 10^{-6}$ & $4.0\cdot 10^{-4}$&   $ 4.0 \cdot 10^{-4} $ \\\hline 
			$r_a=900$, $r_p=750$, ($e=~0.0909$, $p=818.1$)  & $ 3.797 \cdot 10^{-5}  $ &  $5.3\cdot 10^{-5}$ & $1.6\cdot 10^{-5}$ &   $ 2.0 \cdot 10^{-4} $ \\\hline 
			$r_a=90$, $r_p=55$, ($e=~0.2413$, $p=68.275$)  & $ 1.6012 \cdot 10^{-3} $  & $1.7 \cdot 10^{-2}$ &$1.6 \cdot 10^{-2}$ &   $ 1.0 \cdot 10^{-2} $ \\\hline  $r_a=60$, $r_p=20$  ($e=~0.5$, $p=30$)  & $  5.4077 \cdot 10^{-3} $& $3.7\cdot 10^{-2}$ & $0.05$ &   $ 0.013 $ \\\hline 
			$r_a=40$, $r_p=10$, ($e=~0.6$, $p=16$)  & $  1.5291 \cdot 10^{-2} $& $0.1$ & $0.11 $ &   $  1 \cdot 10^{-2} $ \\\hline 
			$r_a=20$, $r_p=5$, \vskip1pt ($e=~0.6$, $p=8$)  & $  5.2386  \cdot 10^{-2} $& $0.3$ & $0.29$ &   $ 0.03 $ \\\hline 
			
\end{tabular}
\end{center}
		\caption{Shift of the longitude of the node per revolution due to gravitomagnetism. The first column defines the orbital parameters and the second gives the node shift calculated from our approach $\Delta_{\Omega}$ (\ref{eq::delta:omega}) in radians. The last three columns give the relative errors as compared to the post-Newtonian result $\Delta_{\Omega}^{PN}$ (\ref{eq::delta::omega::PN}) and the Kerr result $\Delta_{\Omega}^{FH}$ (\ref{eq::kerr::Omega::FH}). Here the spin parameter is $a=10^{-1}$ and Schwarzschild radius is $r_g=1$.  As initial conditions we choose $\varphi_0 = 0$. Note that the result does not depend on the initial values of the inclination or the longitude of the ascending node.
 } 
	
\label{tab:my_label_1_plus_2}
	\end{table}

	\begin{table}
		\centering
             \begin{center}
 \begin{tabular}{ | m{4.6cm} | m{2.7cm}| m{2.cm} | m{2.cm}|  m{2.cm}|  } 
  \hline
			& $\Delta_{v}    $ &  $|1-\frac{\Delta_{v}}{\Delta_{v}^{PN}}|$   & $|1-\frac{\Delta_{v}^{FH}}{\Delta_{v}^{PN}} |$  
			& $|1-\frac{\Delta_{v}^{FH}}{\Delta_{v}}|$ 
			\\\hline
			$r_a=2\cdot10^{ 5}$, $r_p=1.5\cdot10^{ 5}$,  ($e=0.14$, $p=~1.71~\cdot10^{ 5}$) & $ -2.0296 \cdot 10^{-10} $& $2.5\cdot 10^{-5}$ & $2.0\cdot 10^{-5}$ &   $ 4.8 \cdot 10^{-6} $ \\\hline 
			$r_a=9000$, $r_p=8500$, ($e=~0.028$, $p=8743$) &  $  -1.7628 \cdot 10^{-8} $ &  $5.1\cdot 10^{-4}$ & $4.0\cdot 10^{-4}$&   $ 1.0 \cdot 10^{-4} $ \\\hline 
			$r_a=900$, $r_p=750$, ($e=~0.0909$, $p=818.1$)  & $ -6.1873 \cdot 10^{-7} $ &  $5.3\cdot 10^{-3}$ & $4.2\cdot 10^{-3}$ &   $ 1 \cdot 10^{-3} $ \\\hline 
			$r_a=90$, $r_p=55$, ($e=~0.2413$, $p=68.275$)  & $-2.7180 \cdot 10^{-5}$  & $6.0 \cdot 10^{-2}$ & $5.1 \cdot 10^{-2}$ &   $ 9.7 \cdot 10^{-3} $ \\\hline  $r_a=60$, $r_p=20$  ($e=~0.5$, $p=30$)  & $ -1.0052\cdot 10^{-4}  $& $0.12 $ & $0.12$ &   $ 1.03 \cdot 10^{-2} $ \\\hline 
			$r_a=40$, $r_p=10$, ($e=~0.6$, $p=16$)  & $ -2.9313 \cdot 10^{-3}  $& $0.23$ & $0.22$ &   $9.2 \cdot 10^{-3} $ \\\hline 
			$r_a=20$, $r_p=5$, \vskip1pt ($e=~0.6$, $p=8$)  & $ -1.1661 \cdot 10^{-3} $& $0.45$ & $0.45$ &   $ 6.1 \cdot 10^{-3} $ \\\hline 
			
\end{tabular}
 \end{center}
        	\caption{Shift of the pericentre per revolution obtained from our approach $\Delta_{v}$ (\ref{eq::delta:phi}), and from the post-Newtonian calculation $\Delta_{v}^{PN}$ (\ref{eq::delta::phi::PN}), together with their ratio, for the Kerr space-time treated as a perturbation of the Schwarzschild background.  We set $r_g=1$, $a=10^{-3}$ and use the initial :  $\iota=1$,  $\varphi_0=0$.}

		\label{tab:my_labeL_z}
	\end{table}


\subsection{Equations for angular momentum $L_z$, energy $E$ and argument of pericentre $\varphi_0$}
Putting the definitions of the force components (\ref{def::fth::kerr})-(\ref{def::fph::kerr}) into equations (\ref{eqv::dot_c3})-(\ref{eqv::dot_c2}) we have \begin{eqnarray}
	\dot{L}_z = - J  r_g^2 r^2  \sin^2\theta    \left(\frac{\dot{t}\dot{r}}{r^4 (1-\frac{r_g}{r})}-\frac{2 \dot{t}\dot{\theta}  \cot \theta }{r^3}\right) , \\
	\dot{E}  = \left(1- \frac{r_g}{r}\right) \frac{3 J  r_g^2 \dot{r} \dot{\phi}  \sin ^2\theta }{r^2 (1-\frac{r_g}{r})} =  \frac{3 J  r_g^2 \dot{r} L_z   }{r^4 } ,
\end{eqnarray}
which can be easily integrated as  
\begin{eqnarray}
	L_z  = L_z(0) + \frac{J  r_g^2  \sin ^2\theta}{r} \dot{t},\label{eq::Delta_L_z}
\end{eqnarray}
and 
\begin{eqnarray}
	 E  = E(0) +  \frac{J  r_g^2  L_z  }{r^3 }.\label{eq::Delta_E}
\end{eqnarray}
Comparing the last two expressions with the conserved quantities defined by the Killing vectors in the Kerr space-time (see, e.g., \cite{cieslik2023kerr}), linearized in $J$,
\begin{eqnarray}
l_z &= r^2 \dot{\phi} \sin^2\theta + \frac{J r_g^2 \sin^2\theta}{r} \dot{t}, \
\epsilon &= \frac{\dot{t}}{1-\frac{r_g}{r}} + \frac{J r_g^2 l_z}{r^3},
\end{eqnarray}
we see that, for $E(0)=\epsilon$ and $L(0)=-l_z$, our solutions (\ref{eq::Delta_L_z}) and (\ref{eq::Delta_E}) correctly reproduce these relations.  Therefore, we can define 
\begin{eqnarray}
	\bar{l}_z  = L_z  - \frac{J r_g^2 \sin ^2\theta}{r} \dot{t}
\end{eqnarray} 
and
\begin{eqnarray}
	\bar{\epsilon}  = E -  \frac{J  r_g^2  l_z  }{r^3 } ,
\end{eqnarray}
as two new osculating elements  with perturbation equations as
\begin{eqnarray}
	\dot{\bar{\epsilon}}  = 0 , \\
	\dot{\bar{l}}_z  = 0,
\end{eqnarray}
which are effectively constants.  

Finally, from equation (\ref{eqv::dot_phib0}) we can define the secular correction for $\bar{\varphi}_0$ as
\begin{eqnarray}
	\bar{\varphi}_0 = \bar{\varphi}_0(s_0) +  (s-s_0)   F_{\bar{\varphi}_0,0},  
\end{eqnarray}
from which we can obtain the shift of pericentre per one revolution (a factor $\frac{1}{2}$ appears in this formula due to our choice of the argument in the radial solution (\ref{eq:solr}))
\begin{eqnarray}
	\Delta_{v}= \frac{1}{2} Re\Delta_{\bar{\varphi}_0} =\frac{s_p}{2}   ReF_{\bar{\varphi}_0,0}\label{eq::delta:phi}
\end{eqnarray}
where we neglect the impact of shift of Weierstrass half-periods $\omega_1$ and $\omega_3$ as the corresponding corrections to 
energy and angular momentum are purely periodic and therefore do not contribute secularly.  Where $F_{\bar{\varphi}_0,0}$ is obtained from the evolution equation (\ref{eqv::dot_phib0}); due to its cumbersomeness, the explicit expression is provided in (\ref{eq::delta:phi::Fbar}).

\begin{figure}
	\centering
	\begin{subfigure}{.48\textwidth}
		\centering
		\includegraphics[width=.99\linewidth]{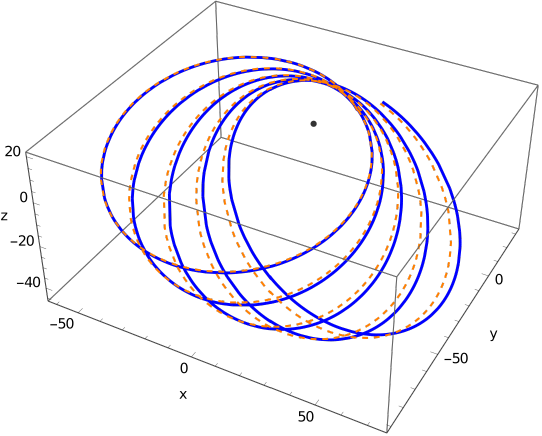}
		\caption{$r_a= 90, r_p=10, a=\frac{1}{10}$}
		\label{fig:trajectoriesrp20ra90a1n10}
	\end{subfigure}
	\hskip0.2cm
	\begin{subfigure}{.48\textwidth}
		\centering
		\includegraphics[width=.99\linewidth]{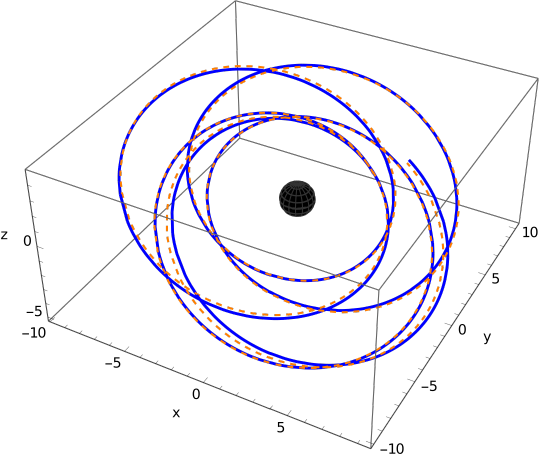}
		\caption{$r_a= 10, r_p=5, a=\frac{1}{100}$}
		\label{fig:trajectoriesrp5ra10a1n100}
	\end{subfigure}%
        \caption{ Comparison of orbital trajectories under the influence of the rotation of the central object. The dashed curves represent the unperturbed Schwarzschild geodesic, while the solid curves show the trajectories modified by secular perturbations.  We set $r_g=1$ (shown as a black sphere in the plot) together with the initial condition: $\iota=\frac{1}{2}$, $\Omega=1$,  $\varphi_0=0 $.} 

	\label{fig:trajectories}
\end{figure}

Using approach similar to the previous section and exact expresion for $F_{\bar{\varphi}_0,0}$ (\ref{eq::delta:phi::Fbar})  we can find that in the post-Newtonian limit
\begin{eqnarray}
	\Delta^{PN}_{v}= -3  \cos \iota \Delta^{PN}_{\Omega} \label{eq::delta::phi::PN}
\end{eqnarray}
which is identical to Lense--Thirring's  original results \cite{lense1918einfluss}, \cite{ruffini2003nonlinear}, \cite{soffel2019applied}. 
In Fig.~\ref{fig:trajectories}, we compare geodesic motion with perturbed trajectories with secular corrections for different initial conditions.

\subsection{Comparison with Kerr geodesics and with post-Newtonian results} \label{sec:compKerr}
\begin{figure}
	\centering
	\begin{subfigure}{.48\textwidth}
		\centering
		
		\includegraphics[width=.99\linewidth]{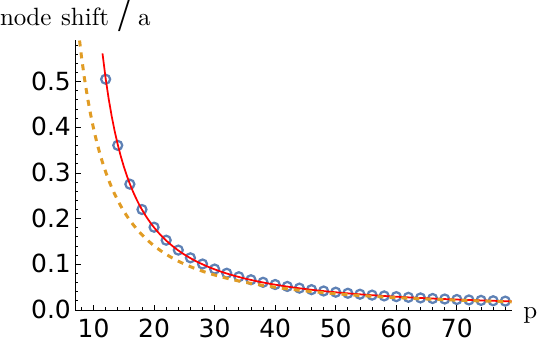}
		\caption{$e=\frac{8}{9}$}
		\label{fig:delta_omega_p_FH_comper_num_e89_FH_p10_80}
		
	\end{subfigure}
	\hskip0.2cm
	\begin{subfigure}{.48\textwidth}
		\centering
		
		\includegraphics[width=.99\linewidth]{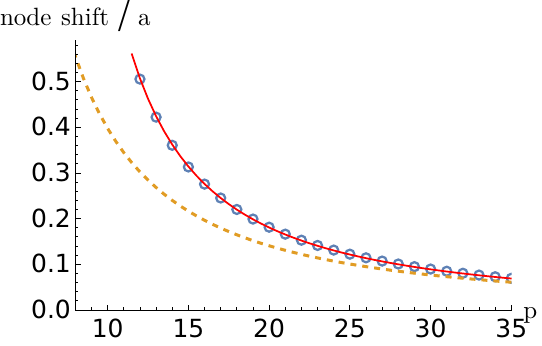}
		\caption{$e= \frac{5}{9}$}
		\label{fig:kerr_delta_omega_comperHL1}
	\end{subfigure}%
    \caption{Comparison between the shift of the longitude of the node per revolution given by our approach $\Delta_{\Omega}$ (\ref{eq::delta:omega}) (red line), the exact analytical result $\Delta^{FH}_{\Omega}$ (\ref{eq::kerr::Omega::FH}) (blue circles), and the first-order post-Newtonian expression $\Delta^{PN}_{\Omega}$ (\ref{eq::delta::omega::PN}) (yellow dashed line), shown as functions of the semi-latus rectum $p$, for different values of eccentricity $e$, for the Kerr space-time treated as a perturbation of the Schwarzschild background. We set   $r_g=2$, $a=10^{-3}$. Note that this result does not depend on the initial value of the inclination.}

\label{fig:delta_p}
\end{figure}

\begin{figure}
\centering
\begin{subfigure}{.48\textwidth}
	\centering
	
	\includegraphics[width=.99\linewidth]{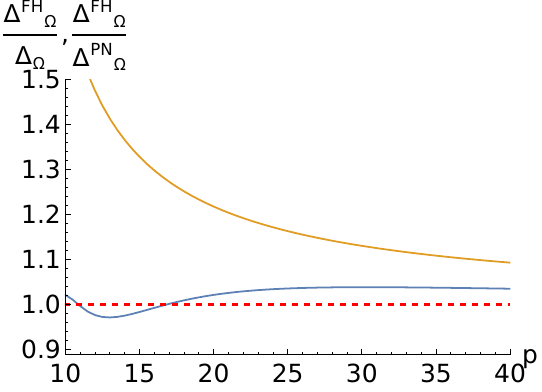}
	\caption{$e=\frac{4}{9}$}
	\label{fig: }
	
\end{subfigure}
\hskip0.2cm
\begin{subfigure}{.48\textwidth}
	\centering
	
	\includegraphics[width=.99\linewidth]{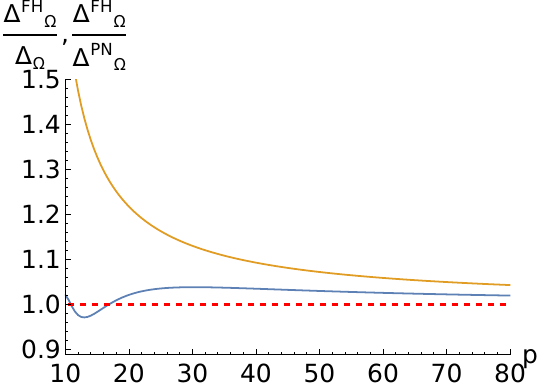}
	\caption{$e= \frac{4}{9}$}
	\label{fig: }
\end{subfigure}%

\centering
\begin{subfigure}{.48\textwidth}
\centering

\includegraphics[width=.99\linewidth]{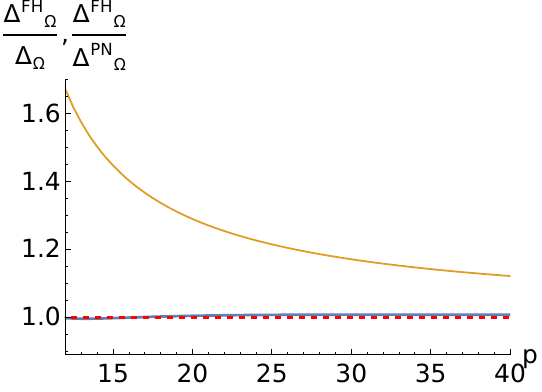}
\caption{$e=\frac{8}{9}$}
\label{fig: }

\end{subfigure}
\hskip0.2cm
\begin{subfigure}{.48\textwidth}
\centering

\includegraphics[width=.99\linewidth]{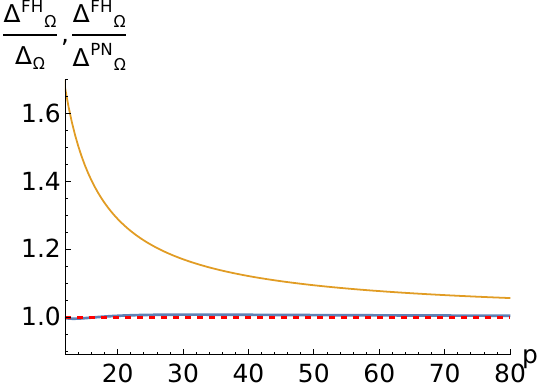}
\caption{$e= \frac{8}{9}$}
\label{fig: }
\end{subfigure}%
\caption{ Comparison between the shift of the longitude of the node per revolution obtained from our approach $\Delta_{\Omega}$ (\ref{eq::delta:omega}), the exact analytical result $\Delta^{FH}_{\Omega}$ (\ref{eq::kerr::Omega::FH}), and the first-order post-Newtonian expression $\Delta^{PN}_{\Omega}$ (\ref{eq::delta::omega::PN}), shown as functions of the semi-latus rectum $p$, for different values of eccentricity $e$, for the Kerr space-time treated as a perturbation of the Schwarzschild background. The dashed red line is the reference unity line, the solid yellow and blue lines represent the ratios $\Delta^{FH}_{\Omega}/\Delta^{PN}_{\Omega}$ and $\Delta^{FH}_{\Omega}/\Delta_{\Omega}$, respectively.    We set  $r_g=2$, $a=10^{-3}$. Note that this result does not depend on the initial value of the inclination. }

\label{fig:delta_comper_e_89}
\end{figure}

In this subsection we compare our results on the secular effects against two useful reference cases. Let us start with the shift of the node. An exact result in Kerr spacetime was derived in  
\cite{fujita2009analytical},  
\begin{eqnarray}
	\Delta^{FH}_{\Omega} =2 \pi \frac{\Upsilon_\phi - \Upsilon_\theta}{\Upsilon_r} ,\label{eq::kerr::Omega::FH}
\end{eqnarray}
where exact expressions for orbital frequencies $\Upsilon_\phi$, $\Upsilon_\theta$, $\Upsilon_r$ are written in terms of elliptic functions \cite{fujita2009analytical}. From table \ref{tab:my_label_1_plus_2} we see that at large distances, as they appear e.g. in the Solar System, the post-Newtonian results are indistinguishable from both the exact solution and our results. However, closer to the BH horizon the post-Newtonian results begin to deviate from the exact solution. As shown in table \ref{tab:my_label_1_plus_2}, and especially in figures~\ref{fig:delta_p} and~\ref{fig:delta_comper_e_89}, our solution (\ref{eq::delta:omega}) provides a very good approximation to the exact results (\ref{eq::kerr::Omega::FH}) even very close to the black hole, and for high eccentricities as well. If we approach the regime of the ISCO, at about $r = 5 r_g$ the error is still on the order of a few percent.

\begin{figure}[h!]
	\centering
	\begin{subfigure}{.48\textwidth}
		\centering
		\includegraphics[width=.99\linewidth]{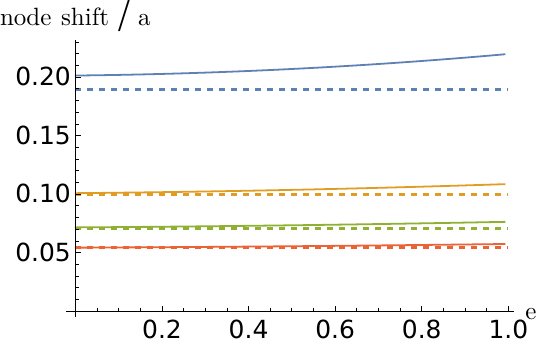}
		\caption{$p \in \left\{13, 20, 25, 30\right\}$}
		\label{fig:kerr_delta_omega_e}
	\end{subfigure}
	\hskip0.2cm
	\begin{subfigure}{.48\textwidth}
		\centering
		\includegraphics[width=.99\linewidth]{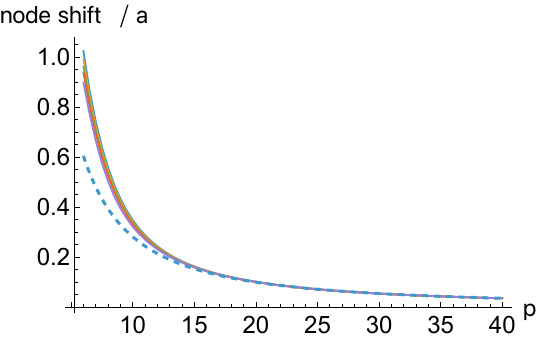}
		\caption{$e\in \left\{\frac{8}{9},\frac{7}{9},\frac{6}{9},\frac{5}{9}, \frac{1}{1000} \right\}$}
		\label{fig:kerr_delta_omega_p}
	\end{subfigure}
    %
    \caption{The shift of the longitude of the node per revolution. Solid lines correspond to our result $\Delta_{\Omega}$ (\ref{eq::delta:omega}) while dashed lines represent the post-Newtonian expression  $\Delta^{PN}_{\Omega}$ \ref{eq::delta::omega::PN}, shown as functions of the eccentricity $e$ (figure~\ref{fig:kerr_delta_omega_e}) and of the semi-latus rectum $p$ (figure~\ref{fig:kerr_delta_omega_p}), for the Kerr space-time treated as a perturbation of the Schwarzschild background. The curves are shown in blue, yellow, green, and red, following the order indicated in the set. We set $r_g=1$ and use the initial condition $\varphi_0=\tfrac{9\pi}{5}$. Note that this result does not depend on the initial value of the inclination.}
    \label{fig:delta_omega}

\end{figure}

From figure~\ref{fig:delta_omega}, we see that, in contrast to the post-Newtonian result (\ref{eq::delta::omega::PN}), our shifts (\ref{eq::delta:omega}) depend on the initial value of the eccentricity.  However, this dependence is negligible for large semi-latus rectum and becomes significant only very close to the BH horizon.

\begin{figure}[h!]
	\centering
	\begin{subfigure}{.48\textwidth}
		\centering
		\includegraphics[width=.99\linewidth]{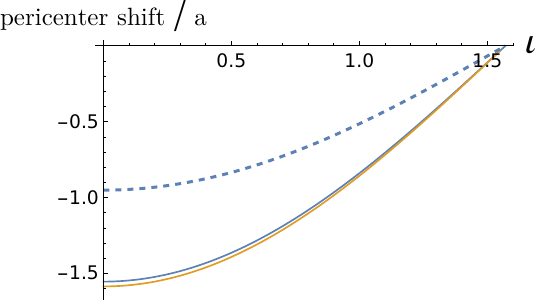}
		\caption{$r_a=20   , r_p= 6 $}
		\label{fig:kerr_delta_phi_e}
	\end{subfigure}
	\hskip0.2cm
	\begin{subfigure}{.48\textwidth}
		\centering
		\includegraphics[width=.99\linewidth]{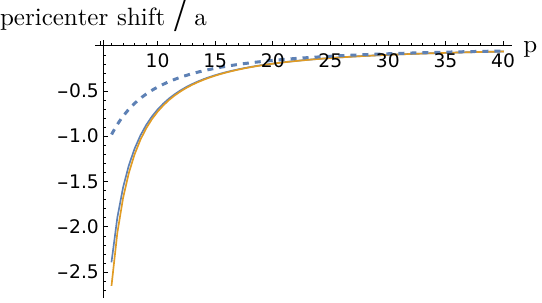}
		\caption{$\iota= 1$}
		\label{fig:kerr_delta_phi_p}
	\end{subfigure}%
        \caption{ The pericentre shift per revolution,  solid blue lines correspond to our result $\Delta_{v}$ (\ref{eq::delta:phi}), solid yellow lines to  the exact analytical result $\Delta^{FH}_{v}$(\ref{eq::kerr::Omega::FH}), while dashed lines represent the post-Newtonian expression $\Delta_{v}^{PN}$ (\ref{eq::delta::phi::PN}), as functions of the inclination $\iota$ (figure~\ref{fig:kerr_delta_phi_e}) and of the semi-latus rectum $p$ (figure~\ref{fig:kerr_delta_phi_p}), for the Kerr space-time treated as a perturbation of the Schwarzschild background. The curves are shown in blue, yellow, green, and red, following the order indicated in the set. We set $r_g=1$ and $a= 10^{-6} $ together with the initial condition $\varphi_0=0 $. }

	\label{fig:delta_phi}
\end{figure}

Next we turn to the shift of the pericentre,
for which we can construct exact analytical expression as
\begin{eqnarray}
	\Delta^{FH}_{v} = 2 \pi \frac{ \Upsilon_\theta}{\Upsilon_r} - 4\omega_1,\label{eq::kerr::v::FH}
\end{eqnarray}
where exact expressions for orbital frequencies $\Upsilon_\phi$, $\Upsilon_r$ are written in terms of elliptic functions \cite{fujita2009analytical} and $4\omega_1$ is a period of Weierstrass elliptic function.
In table~\ref{tab:my_labeL_z}, we compare our results with the post-Newtonian approximation and the exact pericentre shift per orbital revolution. We find that at large orbital radii the post-Newtonian results are practically indistinguishable from both the exact solution and our approach. However, closer to the black hole horizon, our results converge significantly faster to the exact solution. Tables~\ref{tab:my_label_1_plus_2} and~\ref{tab:my_labeL_z} further show that the post-Newtonian approximation deviates from our solution more rapidly for the pericentre shift than for the nodal precession. Finally, in figure~\ref{fig:delta_phi} we compare our results~(\ref{eq::delta:phi}) with the post-Newtonian expression~(\ref{eq::delta::phi::PN}) and the exact solution as functions of the inclination $\iota$ and the semi-latus rectum $p$.

\section{ Orbital perturbations from quadrupoles}	\label{qmetr::sol}

The deviation of the central object from spherical symmetry is one of the most important sources of a perturbation acting on orbiting objects. This deviation can be modeled by a nonvanishing quadrupole moment of the gravitational field, as e.g. given by the q-metric. 
This metric, also known as Zipoy-Voorhees metric or $\gamma$-metric, is a static and axially symmetric solution of the vacuum Einstein equation. It possesses a naked singularity \cite{quevedo2011quadrupolar} (which might be covered by an interior solution) and can be interpreted as describing the external gravitational field of a mass with a quadrupole moment. Its astrophysical relevance has been investigated by many authors, see e.g. \cite{faraji2022circular,idrissov2025geodesic,arrieta2020shadows,boshkayev2021luminosity,momynov2024gravitational,lora2023q,allahyari2019quasinormal,katsumata2025periapsis} and references therein. We will utilize the q-metric here to derive a perturbation force for our scheme. Note that our approach here may also help to better understand geodesic motion in the quadrupolar space-time \cite{herrera2000geodesics,boshkayev2016motion}. 

As shown in \cite{allahyari2019quasinormal}, to interpret $q$ as the quadrupole parameter in the post-Newtonian expansion it is necessary to introduce an additional coordinate system. We derive relations between the osculating elements in our coordinate system and those in the additional one in \ref{Appendice::q_metr_interpretation}.       

In a Schwarzschild-like coordinate system we can write the $q$-metric as
\begin{eqnarray}
&g_{tt}=   \left(1- \frac{r_g}{r}\right)^{1+q} ,\\
&g_{rr}= -\left(1- \frac{r_g}{r}\right)^{-q-1} \left(1+ \frac{r_g^2 }{4 r^2 }\frac{\sin^2\theta }{1-r_g/ r}\right)^{-q(2+q)},\\
&g _{\theta\theta}=- \left(1- \frac{r_g}{r}\right)^{-q} \left(1+ \frac{r_g^2}{4 r^2}\frac{ \sin^2\theta }{1-r_g/r }\right)^{-q(2+q)} r^2,\\
&g_{\phi\phi}=-\left(1- \frac{r_g}{r}\right)^{-q}r^2 \sin^2 \theta,
\end{eqnarray}
where $r_g=2m$ and $q$ is a dimensionless parameter related to the mass quadrupole moment of the source as $ M_1 = -\frac{m^3}{3} q (1+q)(2+q) $, and to the ADM mass as $M_0=(1+q)m$ \cite{quevedo2011quadrupolar}. Writing geodesic equations for this metric and expanding them in a series for small $q$, we can define an external force for the Schwarzschild space-time in linear order as
\begin{eqnarray}
f^t = -q  \frac{ \dot{t} \dot{r} r_g}{r^2(1- \frac{r_g}{r})} , \\
f^\theta =  q \left(  \dot{\phi}^2 f_{\phi,\phi}^{\theta}+\dot{\theta}  \dot{r}f_{\theta,r}^{\theta}+  \dot{\theta}^2 f_{\theta,\theta}^{\theta} + \dot{r}^2 f_{r,r}^{\theta}  \right),\label{def::fth::q}\\
f^\phi =  q  \frac{ \dot{\phi} \dot{r} r_g}{r^2 (1-\frac{r_g}{r})} \label{def::fph::q} ,
\end{eqnarray}
where
\begin{eqnarray}
f_{\phi,\phi}^{\theta} = \sin 2 \theta   \ln \left(1+\frac{r_g^2 }{4 r^2 }\frac{\sin^2\theta }{1-r_g/ r} \right),\\
f_{\theta,r}^{\theta}=\frac{ r_g \csc ^2\theta   \left(8 r^2+r_g \cos 2 \theta  (4 r-3 r_g)-12 r r_g+3 r_g^2\right)}{2 r (r-r_g) \left(4 r \csc ^2\theta  (r-r_g)+r_g^2\right)},\\
f_{\theta,\theta}^{\theta}=\frac{ r_g^2 \sin 2 \theta  }{4 r (r-r_g)+r_g^2 \sin ^2\theta },\\
f_{r,r}^{\theta} = -\frac{2 r_g^2 \cot \theta  }{r (r-r_g) \left(4 r \csc ^2\theta  (r-r_g)+r_g^2\right)}.
\end{eqnarray}
Note that the expressions for $f^t$ and $f^\phi$ are actually exact, and higher order in $q$ do not appear. This will simplify some calculations later on.

\subsection{Equations for angular momentum $L_z$ and energy $E$}
We can write equation (\ref{eqv::dot_c2}) for $E$ as
\begin{eqnarray}
&\dot{E}= -q  \frac{ E \dot{r} r_g}{r^2(1- \frac{r_g}{r})} ,
\end{eqnarray}
which we can easily integrate 
\begin{eqnarray}
& E =  E^0  \left(1-\frac{r_g}{r}\right)^{-q},
\end{eqnarray}
where $E^0$ is a constant of integration. It can be calculated by using the osculating condition, which implies that $E$ has the same form as for geodesics in Schwarzschild spacetime, resulting in $E^0 = \dot{t} (1-\frac{r_g}{r})^{q+1}$. Comparing with known results \cite{herrera2000geodesics}, we see that $E^0$ exactly reproduces the constant energy along geodesics in the q-metric, $E^0=E_{q}=\rm const$.

Analogously, we have equation (\ref{eqv::dot_c3}) for $L_z$
\begin{eqnarray}
&\dot{L}_{z}=  q \frac{L_z  \dot{r} r_g}{r^2 (1-\frac{r_g}{r})} ,
\end{eqnarray}
which is easy to integrate as
\begin{eqnarray}
& L_z= L_z^0\left(1-\frac{r_g}{r}\right)^{q}   ,
\end{eqnarray}
where $L_z^0$ is again a constant of integration. By the same argument, we also reproduce the angular momentum from \cite{herrera2000geodesics} as $L_z^0=-L_{q}= \dot{\phi} \; r^2 \sin^2  \theta \; (1-\frac{r_g}{r})^{-q}$.

 To use these formulas for the description of motion, we first linearize them in $q$.  In the linear approximation in $q$, we have 
\begin{eqnarray}
& E= E^0 - q \; E^0  \ln{\left(1-\frac{r_g}{r}\right)},
\end{eqnarray}
and
\begin{eqnarray}
& L_z= L_z^0 + q \; L_z^0  \ln{\left(1-\frac{r_g}{r}\right)}.
\end{eqnarray}

\subsection{Equations for inclination $\iota$ and longitude of the node $\Omega$ }
Putting the definitions of the force components (\ref{def::fth::q})-(\ref{def::fph::q}) into equations (\ref{eq::diota})-(\ref{eq::domega}) for inclination $\iota $ and longitude of the node $\Omega$ we have 
\begin{eqnarray}
\dot{\iota} &= -q\frac{ r^2 \cot \varphi  \sin ( \Omega -\phi)}{L}  \left(  \dot{\phi}^2 f_{\phi,\phi}^{\theta}+\dot{\theta}  \dot{r}f_{\theta,r}^{\theta} +  \dot{\theta}^2  f_{\theta,\theta}^{\theta} + \dot{r}^2 f_{r,r}^{\theta}  \right) \nonumber
\\ &+   q \frac{ \sin \iota  \cos \iota \cos ^2\varphi   }{\sin^2 \theta} \frac{  \dot{r} r_g}{r^2  (1-\frac{r_g}{r})} ,\label{eq::q_metr::delta::iota} \\
\dot{\Omega} &=q \frac{ r^2 \sin (\Omega -\phi)}{L \sin\iota }   \left(  \dot{\phi}^2 f_{\phi,\phi}^{\theta}+\dot{\theta}  \dot{r}f_{\theta,r}^{\theta}+  \dot{\theta}^2 f_{\theta,\theta}^{\theta} + \dot{r}^2 f_{r,r}^{\theta}  \right) \nonumber
\\ & -q \frac{ \sin \varphi  \cos\varphi}{\sin^2 \theta}    \frac{  \dot{r} r_g}{r^2 (1-\frac{r_g}{r})} \label{eq::q_metr::delta::omega} .
\end{eqnarray}
From these equations we can calculate the secular shifts of the node and the inclination as defined in (\ref{eq:def::sec:cor}),
\begin{eqnarray}		 &\Delta_{\Omega}^{q} =s_p  F_{\Omega,0} , \label{eq::q_metr::delta::omega::sol}  \end{eqnarray} 
and
\begin{eqnarray}		 &\Delta_{\iota}^{q} = s_p    F_{\iota,0}\,, \label{eq::q_metr::delta::i}  \end{eqnarray} 
where $s_p$ is again the proper orbital period and $F_{X,0}$ we indicate the zeroth harmonics of the osculating element $X$. In the simplest case of $\iota=0$,  for longitude of the node $\Omega$ we have 
\begin{eqnarray}		 &\Delta_{\Omega}^{q} =q\int_0^{4\omega_1}  \sin^2\varphi \frac{   r^3 \wp'^2 - 
4\left(r - r_g \right)\left(r_g - 
2 r \right)^2\log\left (1+ \frac{r_g^2} {4 r^2 - 
	4 r r_g}  \right)  } {2r \left(1 - 
\frac{r_g}{r} \right)\left(2r-r_g \right)^2}  d\varphi \nonumber \\
& +q\int_0^{4\omega_1} \sin\varphi \cos \varphi \frac{  
r_g \wp'  } { \left(1 - 
\frac{r_g}{r} \right)\left(2r-r_g \right)}d\varphi, \label{eq::q_metr::delta::omega::i0}  \end{eqnarray} 
and for inclination $\iota$  
\begin{eqnarray}		 &\Delta_{\iota}^{q} = 0. \label{eq::q_metr::delta::i::i0}  \end{eqnarray} 
In the post-Newtonian limit we have for (\ref{eq::q_metr::delta::omega::sol}) 
\begin{eqnarray}		 &\Delta_{\Omega}^{PN,q} =-q \frac{   \pi  r_g^2  }{2 p^2}\cos\iota  
\label{eq::q_metr::delta::omega::PN}  \end{eqnarray} 
and for (\ref{eq::q_metr::delta::i})  
\begin{eqnarray}		 &\Delta_{\iota}^{PN,q} = 0. \label{eq::q_metr::delta::i::PN}  \end{eqnarray} 

\begin{figure}[h!]
\centering
\begin{subfigure}{.48\textwidth}
\centering
\includegraphics[width=.99\linewidth]{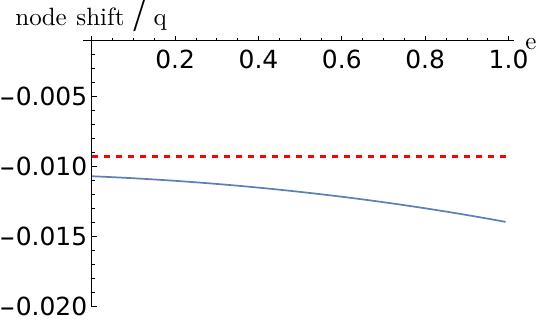}
\caption{$\iota= 0$}
\label{fig:q_metr_delta_omega_e_p_13_i_0}
\end{subfigure}
\hskip0.2cm
\begin{subfigure}{.48\textwidth}
\centering
\includegraphics[width=.99\linewidth]{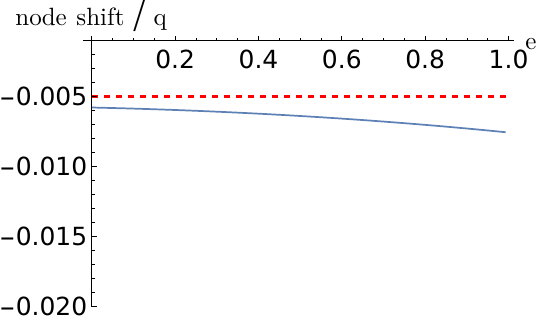}
\caption{$\iota=1$}
\label{fig:q_metr_delta_omega_e_p_13_i_1}
\end{subfigure}%

\caption{The shift of the longitude of the node per revolution as a function of the eccentricity $e$, for different initial values of the inclination, for the quadrupolar space-time treated as a perturbation of the Schwarzschild background. The blue solid lines correspond to our result $\Delta_{\Omega}^q$ (\ref{eq::q_metr::delta::omega::sol}), while the red dashed lines represent the post-Newtonian expression $\Delta_{\Omega}^{PN,q}$ (\ref{eq::q_metr::delta::omega::PN}).  We set $r_g=1$ and use the initial: $p=13$, $\varphi_0=0$.}

\label{fig:q_metr_delta_omega_e}
\end{figure}

Expressions  (\ref{eq::q_metr::delta::omega::PN}) and (\ref{eq::q_metr::delta::i::PN})  reproduce the textbook result for the quadrupole moment correction to the Keplerian orbit (\cite{Klioner}). As we see from figure~\ref{fig:q_metr_delta_omega_e}, it correctly reproduces the behavior of the node shift $\Delta_{\Omega}^{q}$ (\ref{eq::q_metr::delta::omega::sol}) for large semi-latus rectum.

\begin{figure}[h!]
\centering
\begin{subfigure}{.48\textwidth}
\centering
\includegraphics[width=.99\linewidth]{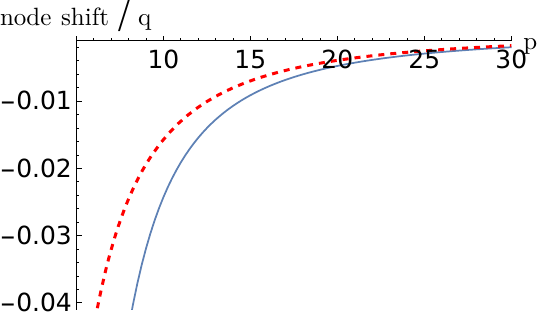}
\caption{$ \iota=0$}
\label{fig:q_metr_delta_omega_p_e_34(i_0)}
\end{subfigure}
\hskip0.2cm
\begin{subfigure}{.48\textwidth}
\centering
\includegraphics[width=.99\linewidth]{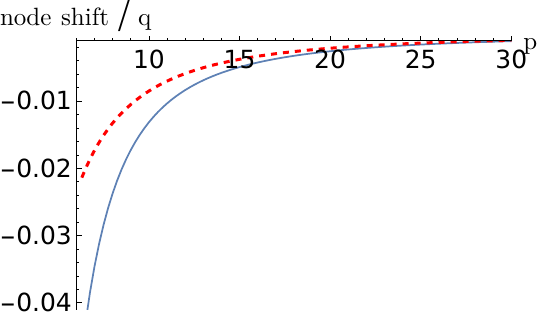}
\caption{$ \iota=1$}
\label{fig:q_metr_delta_omega_p_e_34(i_1)}
\end{subfigure}%

\caption{The shift of the longitude of the node per revolution as a function of the semi-latus rectum $p$, for different initial values of the inclination, for the quadrupolar space-time treated as a perturbation of the Schwarzschild background. The blue solid lines correspond to our result $\Delta_{\Omega}^q$ (\ref{eq::q_metr::delta::omega::sol}), while the red dashed lines represent the post-Newtonian expression $\Delta_{\Omega}^{PN,q}$ (\ref{eq::q_metr::delta::omega::PN}).  We set $r_g=1$ and use the initial:  $e=\frac{3}{4}$, $\varphi_0=0$.}
\label{fig:q_metr_delta_omega_p}
\end{figure}

\begin{figure}[h!]
\centering
\begin{subfigure}{.48\textwidth}
\centering
\includegraphics[width=.99\linewidth]{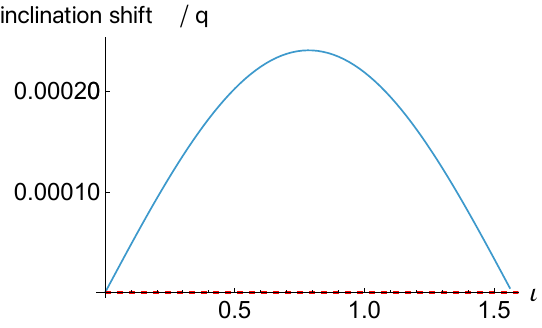}
\caption{}
\label{fig:q_metr_delta_i_i_10_20}
\end{subfigure}
\hskip0.2cm
\begin{subfigure}{.48\textwidth}
\centering
\includegraphics[width=.99\linewidth]{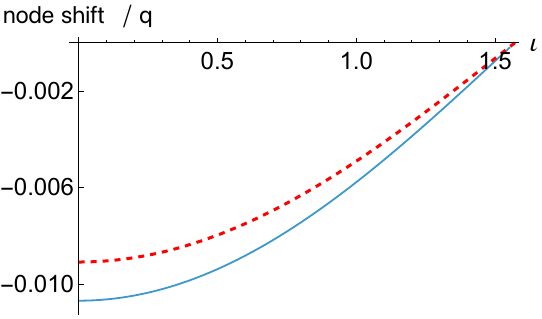}
\caption{}
\label{fig:q_metr_delta_omega_i_10_20}
\end{subfigure}%
\caption{The shifts of the inclination $\Delta_{\iota}^{q}$ (\ref{eq::q_metr::delta::i}) (figure~\ref{fig:q_metr_delta_i_i_10_20}) and of the longitude of the node per revolution $\Delta_{\Omega}^{q}$~(\ref{eq::q_metr::delta::omega::sol}) (figure~\ref{fig:q_metr_delta_omega_i_10_20}), and their post-Newtonian expressions  $\Delta_{\iota}^{PN,q}$ (\ref{eq::q_metr::delta::i::PN}) and $\Delta_{\Omega}^{PN,q}$ (\ref{eq::q_metr::delta::omega::PN}) as functions of the initial inclination $\iota$, for the quadrupolar space-time treated as a perturbation of the Schwarzschild background. The blue solid lines correspond to our results, while the red dashed lines to the post-Newtonian ones.     We set $r_g=1$ and use the initial: $r_a=20$, $r_p=10$, $\varphi_0=0$.}

\label{fig:q_meter_delta_omega}
\end{figure}

Figures~\ref{fig:q_metr_delta_omega_p} and~\ref{fig:q_metr_delta_omega_e} show that, for different initial inclinations, the post-Newtonian asymptotics work well for large semi-latus rectum $p$, and the main difference appears only at relatively small distances from the horizon. Also, from figure~\ref{fig:q_metr_delta_omega_e} we can see that, in the limit of small semi-latus rectum $p$, the node shift begins to depend on the eccentricity. From figure~\ref{fig:q_metr_delta_i_i_10_20}, we see that the inclination shift is at least two orders smaller than the shift of the node, which  agrees with the post-Newtonian asymptotics. Figure~\ref{fig:q_metr_delta_omega_i_10_20} shows that we correctly reproduce the cosine behavior of the node shift, and that the largest relativistic effect occurs for small initial inclination, $\iota = 0$.

It is also possible to formulate an evolution equation for the argument of pericentre; however, due to the considerable algebraic complexity of the resulting expressions and the associated computational challenges, we leave a detailed analysis of this effect to future work.

\newpage

\section{Discussion and conclusion} \label{disc::conc}
In this manuscript we expand the results of our previous paper \cite{yanchyshen2024gaussian}, developing a perturbation technique for osculating elements in the Schwarzschild space-time background in  terms of Weierstrass elliptic functions for arbitrary gravitational and non-gravitational forces. In addition to the previously considered osculating elements such as energy $E$, angular momentum $L_z$ and argument of pericentre $\varphi_0$ we take into account inclination $\iota$ and longitude of the node $\Omega$, together with the anomalies $M_s$ and $M_t$. Using Hagihara’s analytic solution of the geodesic equations, the perturbation equations for the inclination $\iota$ and the longitude of the ascending node $\Omega$ are almost identical to those known in the Newtonian case, in contrast to the results of \cite{warburton2017evolution}, where the authors used the Chandrasekhar $\chi$ parametrisation. 

As in Newtonian case, the denominator of equation (\ref{eqv::dot_phib0}) for the argument of pericentre  proportional to $e^2$, making the expression ill-defined in the limit $e \to 0$. A consistent treatment of this case requires reparametrization (see, e.g., \cite{PP2007}). Geometrically, since all points on a circular orbit are equivalent, the notions of pericenter and apocenter—and hence the pericenter shift—lose their meaning. We therefore omit this limiting case. Numerically, for large semi-latus rectum $p$ (e.g., Solar System orbits), equation (\ref{eqv::dot_phib0}) remains well-behaved even for small $e$, while the problem becomes critical for nearly circular orbits near the black hole horizon.

As an application of our technique we consider motion in the Kerr and quadrupole space-times as perturbations of the Schwarzschild space-time. We solve the evolution equations for the osculating elements $E$, $L_z$, $\Omega$, $\iota$ and $\varphi_0$ in the linear approximation with respect to a small parameter in the perturbing force and compute the corresponding secular corrections.
For the Kerr space-time we compare our solution to the known exact analytical and post-Newtonian solutions. This comparison shows that our method provides more accurate results than the post-Newtonian approach and our results converge to the exact analytical ones for initial conditions close to the black hole horizon.

Using our approach, it is possible to study motion near a slowly rotating BH in different plasma models, to solve the Mathisson–Papapetrou–Dixon equation in various approximations, and to investigate Love numbers corrections and their impact on geodesic motion. Another potential application of our technique could be its implementation as part of a kludge gravitational waveform scheme.


Using the results from \cite{cieslik2023kerr} or \cite{hackmann2010geodesic}, where the geodesic equations for the Kerr space-time were solved in terms of Weierstrass elliptic functions, one could attempt to develop a perturbation technique for the Kerr space-time background. This could be highly useful for various astrophysical applications; however, from a technical point of view, the development of such methods remains a challenging problem.

\section*{Acknowledgments} 

 We thank Daryna Bukatova and Niels Warburton  for helpful discussions. Support by the Deutsche Forschungsgemeinschaft (DFG, German Research Foundation) under Germany’s Excellence Strategy-EXC-2123 “QuantumFrontiers” -- grant no. 390837967, the Research Unit FOR 5456 "Time as Observable in Geodesy", and the project "General relativistic theory of charged accretion
disk structures around black holes" -- grant no. 510727404 is gratefully acknowledged. O.Y. thanks University College Dublin for the kind hospitality during the completion of this work. This publication has emanated from research conducted with the financial support of Research Ireland under grant number 22/RS-URF-R/3825 (S1).

\newpage

\appendix

\section{Newtonian limit} \label{appendix::newtonian::limit}	
In order to obtain the Newtonian limit for  the radial solution (\ref{eq:solr}) we can use the half-period addition formula  \cite{WW}
\begin{equation}\label{wp_omega_represent_wp}
  \wp(u+ \omega_3) = e_3 +\frac{(e_3-e_1)(e_3-e_2)}{\wp(u) -e_3},
\end{equation}
where $e_i$ are the roots of $\wp'(x)=0$. Using the definition of $e_i$ as the extrema of the Weierstrass elliptic function $\wp(x)$ and the radial solution (\ref{eq:solr}) together with the relation $e_1+e_2+e_3=0$, the roots can be expressed in terms of the pericentre $r_p$ and apocentre $r_a$ distances as
\begin{align} \label{def::ei_ra_rp}
	e_1 =  \frac{2}{3}-   r_g \frac{ r_a +r_p  }{r_a r_p},\\e_2 =  -\frac{1}{3}+\frac{r_g}{r_p},\\ e_3 = -\frac{1}{3} +\frac{r_g}{r_a},\label{def::e3_p_e}
\end{align}
or, equivalently, in terms of the semi-latus rectum $p$ and the eccentricity $e$:
\begin{align} \label{def::ei_p_e}
	e_1 =  \frac{2}{3}-\frac{2 r_g}{p},\\e_2 =  -\frac{1}{3}+\frac{(e+1) r_g}{p},\\ e_3 = -\frac{1}{3} -\frac{(e-1) r_g}{p}.
\end{align}
Furthermore, for the invariants of the Weierstrass elliptic function we have 
\begin{align} \label{def::g2_rp_ra}
	g_2 &=  \frac{4}{3}-\frac{4 r_g }{r_a} \left(1-\frac{r_g}{r_p}\right)+\frac{4 r_g^2}{r_a^2}+\frac{4 r_g^2}{r_p^2}-\frac{4 r_g}{r_p},\\ 
    g_3 &= \frac{4 \left(r_a-3 r_g\right) \left(3 r_g-r_p\right) \left(3 r_a r_g-2 r_a r_p+3 r_g r_p\right)}{27 r_a^2 r_p^2},
\end{align}
which, in the Newtonian limit, reduce to  
\begin{align} \label{def::g2_newtonian}
	g_2 &=  \frac{4}{3},\\ 
    g_3 &= \frac{8}{27}.\label{def::g3_newtonian}
\end{align}
Using relation (\ref{wp_omega_represent_wp}) and expressions of the roots (\ref{def::ei_ra_rp})-(\ref{def::e3_p_e}) we have
	\begin{equation}
  r =  r_a-\frac{3 r_a \left(r_a-r_p\right) \left(-2 r_p r_g-r_a r_g+r_a r_p\right)}{-3 r_a^2 r_g-3 r_p r_a r_g+3 r_p^2 r_g +3 r_p r_a^2-2 r_p^2 r_a+12 r_p^2 r_a \wp(u)}. 
\end{equation}
In the limit $\tfrac{r_g}{r_p} \to 0$, we use the homogeneity property of the Weierstrass function \cite{WW},
\begin{equation}
\wp(\lambda u, \lambda^{-4} g_2, \lambda^{-6} g_3 ) = \lambda^{-2} \wp(u, g_2,g_3),
\end{equation}
together with its trigonometric representation \cite{FunctionsWolfram}
\begin{equation}
\wp \left(u,3,1\right) = \frac{3}{2}\cot^2 \left(\sqrt{\frac{3}{2}},u\right)+1,
\end{equation}
and the Newtonian values of the invariants $g_2$ and $g_3$ (\ref{def::g2_newtonian}), (\ref{def::g3_newtonian}). This yields
\begin{equation}
\wp\left(u, \tfrac{4}{3}, \tfrac{8}{27}\right) = \cot^2(u) + \tfrac{2}{3}.
\end{equation}
Finally, for the radial coordinate we obtain
\begin{align}
r = \lim_{\tfrac{r_g}{r_p}\to 0}\frac{r_g}{\tfrac{1}{3}+\wp(u+\omega_3)}
= \frac{r_a r_p \csc^2(u)}{r_a+r_p \cot^2(u)}
= \frac{p}{1 - e \cos(2u)} ,
\end{align}
from which, using
$u = \tfrac{1}{2}(\phi - \phi_0) + \omega_1$
and noting that $\omega_1 \to \pi$ in the Newtonian limit, we obtain
\begin{align}
r = \frac{p}{1 + e \cos(\phi - \phi_0)} ,
\end{align}
which coincides with the standard Newtonian expression for a Keplerian ellipse in terms of the semi-latus rectum $p$ and the eccentricity $e$ and the argument of periastron $\phi_0$.

\section{Explicit expressions} \label{exact::expressions::eq} 
The explicit expressions for the coefficients in equation (\ref{eqv::dot_phib0}) for the osculating element $\bar \varphi_0$ are
\begin{eqnarray}
	A_t &=  \left(1-\frac{r_g}{r}\right) \frac{2 E  (3 g_2-4)   }{3 \wp ' \left(g_2^3-27 g_3^2\right)   }      \left(6 g_2 \left(\zeta  \wp '+2 \wp ^2\right)-9 g_3 \left(v \wp '+2 \wp \right)-2 g_2^2\right) ,\label{eqv::dot_phib0::at}~~~~~~~~~~~~\\
	A_\phi&= \frac{r^2}{54 L \left(g_2^3-27 g_3^2\right) \wp '} \left( \left(2592 g_3-24 g_2 \left(27 g_3+16\right)\right) \wp ^2 \right.\nonumber\\& \left.+36 \left(3 g_2^3-4 g_2^2+2 g_3 \left(8-27 g_3\right)\right) \wp +4 g_2 \left(g_2 \left(27 g_3+16\right)-108 g_3\right)
	\right.\nonumber\\& \left. +6 \wp' \left(-2 \zeta  \left(27 g_3+16\right) g_2+6 g_3 \left(36 \zeta -27 g_3 v+8 v\right)+9 g_2^3 v-12 g_2^2 v\right) \right) ,\label{eqv::dot_phib0::phi} \\
	A_\theta&= \frac{r^2}{54  L \left(g_2^3-27 g_3^2\right) \wp '} \left( - \left(384 g_2+648 \left(g_2-4\right) g_3\right) \wp ^2 \right.\nonumber\\& \left.+6  \wp'  \left(-2 \zeta  \left(27 g_3+16\right) g_2+6 g_3 \left(36 \zeta -27 g_3 v+8 v\right)+9 g_2^3 v-12 g_2^2 v\right)\right.\nonumber\\& \left.+36 \left(\left(3 g_2-4\right) g_2^2+2 g_3 \left(8-27 g_3\right)\right) \wp  \right.\nonumber\\& \left.+4 g_2 \left(16 g_2+27 \left(g_2-4\right) g_3\right) \right),\label{eqv::dot_phib0::th}
\end{eqnarray}
where $v = \frac{1}{2} (\varphi+\bar\varphi_0)$ and $\zeta= \zeta(\frac{\varphi-\varphi_0}{2}+\omega_2 )$  is the Weierstrass zeta function.

The explicit expressions for the coefficients in equation (\ref{eqv::dot_phib0}) for the osculating element $\bar \varphi_0$ are
\begin{eqnarray}
	B_r &=  -\frac{L (3 g_2-4) \left(4 g_2^2-12 g_2 \left(\zeta  \wp '+2 \wp ^2\right)+9 g_3 \left(2 v \wp '+4 \wp \right)\right)}{6 r_g \left(g_2^3-27 g_3^2\right)} ,\label{eqv::dot_phib0::br}~~~~~~~~~~~~\\
	B_\phi&= \frac{3 r_g^2}{L \left(g_2^3-27 g_3^2\right) (3 \wp +1)^2}  \left(g_2^2 \left(4 (9 \zeta  \wp +v)+6 \wp '\right)  \right. \nonumber \\& \left.-18 g_3 \left(4 \zeta +3 v \left(\wp '\right)^2-4 v \wp +6 \wp  \wp '\right)  \right. \\& \left. -6 g_2 \left(-6 \zeta  \left(\wp '\right)^2+8 \zeta  \wp -9 g_3 (\zeta -v \wp )+\left(4-12 \wp ^2\right) \wp '\right)-3 g_2^3 v\right) \nonumber   ,\label{eqv::dot_phib0::bphi} \\
	B_\theta&=  - \frac{18 r_g^2  \sin i \cos \varphi   }{L \left(g_2^3-27 g_3^2\right) (3 \wp +1)^2} \left(  \wp '  \left(4 g_2 \left(3 \wp ^2-1\right)-18 g_3 \wp +g_2^2\right)    \right. \nonumber \\& \left.  +  \left(  \zeta  g_2 \left(3 g_3+24 \wp ^3-8 \wp \right)+3 g_3 \left(-4 \left(\zeta +3 v \wp ^3-v \wp \right)\right)+v \frac{(3 g_3 -3 g_2^3  +4 g_2^2) }{6} \right) \right) \nonumber  \\& + \frac{9 r_g^2 \cos i \cot i  \sin \varphi }{L (3 \wp +1)^2} ,\label{eqv::dot_phib0::bth}
\end{eqnarray}
where $v = \frac{1}{2} (\varphi+\bar\varphi_0)$ and $\zeta= \zeta(\frac{\varphi-\varphi_0}{2}+  \omega_2 )$  is the Weierstrass zeta function.
For expression of $F_{\bar{\varphi}_0,0}$ in the orbital perturbations from gravitomagnetism cause from evolution equation (\ref{eqv::dot_phib0}) and definition of force components  (\ref{def::ft::kerr})-(\ref{def::fph::kerr}), after simplification we have 
\begin{eqnarray}
	   F_{\bar{\varphi}_0,0}& = -\frac{1}{s_p} \frac{a  E \cos \iota }{36  L \left(g_2^3-27 g_3^2\right)}  \int_{0}^{4\omega_1}  -\frac{8}{9} (3 \wp +1)^2 \left(9 g_3 (v \wp'+2 \wp )-6 g_2 \left(\zeta \wp'+2 \wp ^2\right)+2 g_2^2\right) \nonumber \\
		 & +\frac{1}{(2-3 \wp )^2} \left(\wp'\left(18 g_3 \left(-4 \zeta +3 g_3 v-12 v \wp ^3+4 v \wp -6 \wp' \wp \right)+g_2^2 (4 v+6 \wp') \right.\right.\nonumber \\
		 & \left.\left.-3 g_2^3 v+6 g_2 \left(3 \zeta  g_3+4 \left(3 \wp ^2-1\right) (2 \zeta  \wp +\wp')\right)\right)\right) d \varphi.\label{eq::delta:phi::Fbar}
\end{eqnarray}

\section{Connection of osculating elements in different coordinate systems }	 \label{Appendice::q_metr_interpretation}
Let us consider the following coordinate transformation \cite{allahyari2019quasinormal}
\begin{align}
\theta =  \vartheta - q \frac{m^2}{\rho^2} \left( 1+ 2 \frac{m}{\rho}+ ...\right) \sin \vartheta \cos\vartheta ,\\
r = \rho \left( 1 - q \frac{m}{\rho} - q \frac{m^2}{\rho^2} \left(1 + \frac{m}{\rho}+ ... \right) \sin^2 \vartheta \right) 
\end{align}
where $r_g= 2 m$, $\rho$ and $\vartheta$ are coordinates in which in the post-Newtonian limit, the quadrupole moment of the source is $Q = \frac{2}{3} q m^3$.

We are looking for a connection between constants of motion in coordinate systems with the old $\theta$, $r$ and the new $\vartheta$, $\rho$ coordinates. First, let us find a relation between $L_z$ and $L'_z$ which are defined as 
\begin{eqnarray}
L'_z = \rho^2 \dot{\phi}  \sin^2{\vartheta}, \;\;\; L_z = r^2 \dot{\phi}  \sin^2{\theta}. \label{eq::app::con::cp3::dphi::1}
\end{eqnarray}
Excluding $\dot{\phi}$ from  (\ref{eq::app::con::cp3::dphi::1})  and expanding $r$ and $\theta$ into power series in $q$, we can express $L_z$ in terms of  $L'_z$   
\begin{eqnarray}
L_z=L'_z-q \frac{L'_z m}{\rho ^3} \left(m^2 \cos (2 \vartheta)+3 m^2+2 m \rho +2 \rho ^2\right).
\end{eqnarray}
Similarly, for inclination and node, comparing relations between $\theta$, $\phi$, $\varphi$ (\ref{eqv::theta_phi_0}) and (\ref{eqv::dot_theta_phi}) for different coordinate systems we have 
\begin{eqnarray}
\iota= \iota'+ q\frac{m^2 \sin \iota'  \left( L_z' (2 m+\rho )-\rho  \cos \iota'  (3 m+\rho ) \sin 2\varphi \dot{\rho} \right)}{L' \rho ^3} ,\\
\Omega= \Omega'  -q  \dot{\rho} \cos \iota'  \frac{2 m^2 (3 m+\rho ) \sin^2 \varphi  }{L' \rho ^2} .
\end{eqnarray}

\section*{References}
\bibliography{bib_GR}
\end{document}